\documentclass[11pt,draftclsnofoot,onecolumn]{IEEEtran} 
\usepackage{times}
\usepackage[final]{graphicx}
\usepackage{amsmath,amsfonts}
\usepackage{amssymb,amsbsy}
\usepackage{times}

\usepackage{cite}
\usepackage{epsf,graphics,epsfig}
\usepackage{subfigure}

\newcommand\ignore[1]{}

\newtheorem{thm}{Theorem}
\newtheorem{lem}{Lemma}
\newtheorem{prop}{Proposition}

\newtheorem{cor}{Corollary}

\newcommand{\bef} { {\sf{bf}} }
\newcommand{\gen}{ {\sf{gen}} }

\newcommand{\trace}{ {\mathrm{Tr}}}
\newcommand{\diag} {{\mathrm{diag}}  }

\newcommand{\opt} { {\sf{opt}} }
\newcommand{\stat} { {\sf{stat}} }

\newcommand{\vertex} { {\sf{vertex}} }

\newcommand{\loc}{  {\sf{loc}} }

\newcommand{\bH} {{\mathbf{H}  } }

\newcommand{\bEe}{{\mathit{E}}}    
\newcommand{\bX} {{\mathbf{X}  } }
\newcommand{\bO} {{\mathbf{O}  } }
\newcommand{\bI} {{\mathbf{I}  } }
\newcommand{\bV} {{\mathbf{V}  } }
\newcommand{\bGamma}{ {\mathbf{\Gamma}} }
\newcommand{\bs} {  {\bf s} }

\newcommand{\perf} { {\sf{perf}} }

\newcommand{\bG}   { {\mathbf{G}}} 
\newcommand{\bD}   { {\mathbf{D}}} 
\newcommand{\bU}   { {\mathbf{U}}} 

\newcommand{\bfLambda}  {{\mathbf{\Lambda}} } 
\newcommand{\bSigma}  {{\mathbf{\Sigma}} }

\newcommand{\St}{ {\mathrm{St}} }
\newcommand{\proj}{  {\mathrm{proj}} } 
\newcommand{\bFa} {  {\bf{V}} } 
\newcommand{\bfa}  { {\mathbf{v}  } }
\newcommand{\target} {  {\sf{target}} }						 
\newcommand{\nulla} {  {\sf{null}} }						 
\newcommand{\glo} { {\sf{rvq}} }
\newcommand{\bC}   { {\mathbf{C}}} 

\newcommand{\bB}   { {\mathbf{B}}}

\newcommand{\bA} {{\mathbf{A}}}

\newcommand{\bu}  { {\mathbf{u}}  }
\newcommand{\by}  { {\mathbf{y}}  }

\newcommand{\bv} { {\mathbf{v}} }

\newcommand{\iid} {  {\sf{iid}} }
\newcommand{\ind} { {\sf{ind}} }

\newcommand{\snr}{{\sf{SNR}} }

\newcommand{\hsp}{\hspace{0.1in} }

\newcommand{\hspp}{\hspace{0.05in} }
\newcommand{\hsppp}{\hspace{0.02in} }

\newcommand{\sinr}{ {\sf{SINR}} } 
\newcommand{\mse} { {\sf{MSE}} }
 
\newcommand{\MMSE}{  {\sf {MMSE}}}
 
\newcommand{\bS}{ {\bf S} }
\newcommand{\bL} {{\bf L}}

\newsavebox{\savepar}

\begin{document}
\title{Quantized Multimode Precoding in Spatially \\ [0 mm] 
Correlated Multi-Antenna Channels} 

\author{\large {\hspace{0.5in}} Vasanthan Raghavan, 
Venugopal V. Veeravalli$^*$, Akbar M. Sayeed 
\thanks{V.\ Raghavan and V.\ V.\ Veeravalli are with the Coordinated Science 
Laboratory and the Department of Electrical and Computer Engineering, University 
of Illinois at Urbana-Champaign, Urbana, IL 61801 USA. A.\ M.\ Sayeed is with the 
Department of Electrical and Computer Engineering, University of Wisconsin-Madison, 
Madison, WI 53706 USA. Email: {\tt{vasanthan\_raghavan@ieee.org, 
vvv@uiuc.edu, akbar@engr.wisc.edu.}} $^*$Corresponding author.} 
\newline 
\thanks{ 
This work was partly supported by the NSF under grant \#CCF-0049089 through the 
University of Illinois, and grant \#CCF-0431088 through the University of 
Wisconsin. 
This paper was presented in part at the 
41st Annual Conference on Information Sciences and Systems, Baltimore, MD, 2007.} }

\maketitle
\vspace{-25mm}

\begin{abstract} 
\noindent 
Multimode precoding, where the number of independent data-streams is adapted 
optimally, can be used to maximize the achievable throughput in multi-antenna 
communication systems. Motivated by standardization efforts embraced by the 
industry, the focus of this work is on systematic precoder design with realistic 
assumptions on the spatial correlation, channel state information (CSI) 
at the transmitter and the receiver, and implementation complexity. For spatial 
correlation of the channel matrix, we assume a general channel model, based on 
physical principles, that has been verified by many recent measurement campaigns. 
We also assume a coherent receiver and knowledge of the spatial statistics at the 
transmitter along with the presence of an ideal, low-rate feedback link from the 
receiver to the transmitter. The reverse link is used for codebook-index 
feedback and the goal of this work is to construct precoder codebooks, adaptable 
in response to the statistical information, such that the achievable throughput 
is significantly enhanced over that of a fixed, non-adaptive, i.i.d.\ codebook 
design. We illustrate how a codebook of semiunitary precoder matrices 
localized around some fixed center on the Grassmann manifold can be skewed in 
response to the spatial correlation via low-complexity maps that can rotate and 
scale submanifolds on the Grassmann manifold. The skewed codebook in combination 
with a low-complexity statistical power allocation scheme is then shown to bridge 
the gap in performance between a perfect CSI benchmark and an i.i.d.\ 
codebook design.

\end{abstract}

\begin{keywords}
\noindent Limited feedback communication, quantized feedback, adaptive coding, 
low-complexity signaling, MIMO systems, channel state information at transmitter, 
precoding, multimode signaling 
\end{keywords}


\section{Introduction}
\label{sec1} 
Research over the last decade has firmly established the utility of multiple 
antennas at the transmitter and the receiver in providing a mechanism to increase 
the reliability of signal reception~\cite{stbc}, or the 
rate of information transfer~\cite{telatar}, 
or a combination of the two. The focus of this work is on maximizing the 
achievable rate under certain communication models that are motivated by wireless systems 
in practice. In particular, we assume a {\em{limited (or quantized) feedback}} 
model~\cite{limfb_honig} with perfect channel state information (CSI) at the receiver, 
perfect statistical knowledge of the channel at the transmitter, and a low-rate 
feedback link from the receiver to the transmitter. 
In this setting, the fundamental problem is to determine the optimal 
signaling$/$feedback scheme that maximizes the average mutual information 
given a statistical description of the channel, signal-to-noise ratio ($\snr$), 
the number of antennas, and the quality of limited feedback. 

A low-complexity approach to 
solving this problem is to first determine the rank of the optimal precoder as a 
function of the statistics, $\snr$, and the quality of feedback. The design of the 
optimal scheme is then, in principle, essentially the same as that of a fixed rank 
limited feedback precoder whose rank is adapted optimally. Motivated by this line 
of reasoning, the main theme of this work is the construction of a systematic, yet 
low-complexity, limited feedback precoding scheme (of a fixed rank) that results in 
significantly improved performance over an open-loop\footnote{There is 
no correlation information at the transmitter in an open-loop scheme. That 
is, the channel is assumed to be i.i.d.\ and an i.i.d.\ codebook design is used.} 
scheme. Towards this goal, we consider a simple block fading$/$narrowband setup where 
spatial correlation is modeled by a mathematically tractable channel 
decomposition~\cite{canonical_jayesh,canonical_bonek,tulino_ind}, and includes as 
special cases the well-studied {\em{i.i.d.}} model~\cite{telatar}, 
the {\emph{separable correlation 
model}}~\cite{chuah}, and the {\emph{virtual representation 
framework}}~\cite{akbar,akbar_and_venu}. Furthermore, we also assume that 
the power-constrained input signals come from some discrete constellation set whereas 
the decoder is assumed to have a simple, linear architecture like 
the minimum mean-squared error (${\sf MMSE}$) receiver. 

While precoding has been studied extensively under the i.i.d.\ 
model~\cite{lee_petersen,salz,yang_roy,scaglione_gia_barb,sampath,sampath2,yang_roy2,scaglione_spbgs,palomar_precode}, 
considerable theoretical gaps exist in the limited feedback setting. The extreme 
case of limited feedback beamforming has been studied 
in the i.i.d.\ setting where the isotropicity\footnote{Isotropic means that 
the dominant right singular vector is equally likely to point along any direction 
in the ambient transmit space. This ambient space of all possible right singular 
vector(s) is referred to as the Grassmann manifold. Precise definitions are 
provided later in the paper.} of the dominant right singular vector 
of the channel can be leveraged to uniformly quantize the space of unit-normed 
beamforming vectors, a problem well-studied in mathematics literature 
as the Grassmannian line 
packing (GLP) problem~\cite{david_grass,kiran_outage}. Alternate constructions based 
on Vector Quantization (VQ)$/$Random Vector Quantization (RVQ) are also 
possible~\cite{honig,bhaskar_rao}. Spatial correlation, however, skews the 
isotropicity of the right singular vector, and hence poses a fundamentally more 
challenging problem. 
While VQ codebooks can be 
constructed for the correlated channel case, the construction suffers from high 
computational complexity and the codebook has to be reconstructed from scratch every 
time the statistics change, thus rendering VQ-type solutions impractical. 
Recently, beamforming codebooks that can be easily adapted to statistical variation 
(with low-complexity transformations) have been 
proposed~\cite{david_corr,xia_corr,raghavan_spl_issue}. The other extreme, limited 
feedback spatial multiplexing, has also been studied~\cite{david_mux,bhaskar_rao2}. 

In the intermediate setting\footnote{Here, $1 < M < \min(N_t, N_r)$ with $N_t$ and 
$N_r$ denoting the transmit and the receive antenna dimensions.} of ${\sf rank}$-$M$ 
precoding, under the i.i.d.\ assumption, the isotropicity property of the dominant 
right singular vector of the channel extends to the subspace spanned by the 
$M$-dominant right singular vectors thereby allowing a Grassmannian subspace 
packing solution~\cite{david_heath_multimode}. In the correlated case, 
the fundamental challenge on how to non-uniformly quantize the space of $M$-dominant 
right singular vectors remains the same as in the beamforming case. 
However, unlike the beamforming case, it is not even 
clear how a codebook designed for 
i.i.d.\ channels can be skewed in response to the correlation. In fact, using an 
i.i.d.\ codebook design in a correlated channel can lead to a dramatic degradation 
in performance (see Figs.~\ref{fig_442kron} and~\ref{fig_443can}).

Our main goal here is to 
construct a systematic semiunitary\footnote{An $N_t \times M$ matrix ${\bf X}$ with 
$M \leq N_t$ is said to be semiunitary if it satisfies ${\bf X}^H {\bf X} = \bI_M$.} 
precoder codebook that is tailored to the spatial correlation, and is easily 
adaptable in response to a change in statistics. The heuristic behind our 
construction comes from our previous study of the asymptotic 
performance of the statistical precoder~\cite{stat_semiunitary}. We 
showed in~\cite{stat_semiunitary} that the performance of the statistical 
precoder is closest to the optimal 
precoder when the number of dominant transmit eigenvalues is equal to the rank 
of the precoder, these dominant eigenvalues are well-conditioned, and the receive 
covariance matrix is also well-conditioned. A channel 
satisfying the above conditioning properties is said to be {\emph{matched}} to the 
communication scheme. Thus, while limited (or even perfect) feedback can only lead to 
marginal performance improvement in {\emph{matched channels}}, in the case of 
{\emph{mismatched channels}} where the relative gap in performance between the statistical 
and the optimal precoders is usually large, the potential benefits of limited feedback 
are more significant. 

This study~\cite{stat_semiunitary} 
suggests that spatial correlation orients the directivity of the 
$M$-dominant right singular vectors of the channel towards the statistically 
dominant subspaces and hence, a non-uniform quantization of the local neighborhood 
around the statistically dominant subspaces is necessary. The realizability of such 
a non-uniform quantization with low-complexity, as well as its adaptability, are 
eased by mathematical 
maps 
that can rotate a root 
codeset (or a submanifold) centered at some arbitrary location on the Grassmann 
manifold ${\cal G}(N_t ,\hsppp M)$ towards an arbitrary center and scale it 
arbitrarily. 

Our design includes a statistical component of dominant $M$-dimensional subspaces 
of the transmit covariance matrix, a component corresponding to local quantization 
around the statistical component, and an RVQ component which can be constructed 
with low-complexity. 
In this context, our construction mirrors and generalizes 
our recent work in the beamforming case~\cite{raghavan_spl_issue}. By 
combining a semiunitary codebook (of a small enough 
cardinality) with a low-complexity power allocation scheme that is related to 
statistical waterfilling, we show via numerical studies that significant performance 
gains can be achieved and the gap to the perfect CSI scheme can be bridged considerably. 

\noindent {\bf{\em{Organization:}}} 
The system setup is introduced in Section~\ref{sec2}. In Section~\ref{sec3}, we 
introduce the notion of mismatched channels where limited feedback precoding 
results in significant performance improvement. 
In Section~\ref{sec4}, limited feedback codebooks that enhance performance 
are proposed and in Section~\ref{sec_sca}, mathematical maps are constructed to 
realize these designs with low-complexity. 
Numerical studies are provided in Section~\ref{sec5} with a 
discussion of our results and conclusions in Section~\ref{sec6}. 

\noindent {\bf \em Notation}: 
The $M$-dimensional identity matrix is denoted by ${\bI}_M$. We use $\bX(i,j)$ and 
$\bX(i)$ to denote the $i,j$-th and $i$-th diagonal entries of a matrix $\bX$. In 
more complicated settings ({\emph{e.g.,}} when the matrix $\bX$ is represented as a 
product or sum of many matrices), we use $\bX_{i,j}$ to denote the $i,j$-th entry. 
The complex conjugate, conjugate transpose, regular transpose and inverse operations 
are denoted by $(\cdot)^{\star}$, $(\cdot)^{H}$, $(\cdot)^{T}$ and $(\cdot)^{-1}$ 
while $\bEe [\cdot]$, $\trace(\cdot)$ and $\det(\cdot)$ stand for 
the expectation, the trace and the determinant operators, respectively. 
The $t$-dimensional complex vector space is denoted by ${\mathbb C}^{t}$. 
We use the ordering $\lambda_1(\bX) \geq \cdots \geq \lambda_n(\bX)$ for the 
eigenvalues of an $n \times n$-dimensional Hermitian matrix $\bX$. The notations 
$\lambda_{\max}(\bX)$ and $\lambda_{\min}(\bX)$ also stand for $\lambda_1(\bX)$ 
and $\lambda_n(\bX)$, respectively.

\section{System Setup} 
\label{sec2} 
We consider a communication model with $N_{t}$ transmit and $N_r$ receive antennas 
where $M$ ($1 \leq M \leq \min(N_t,N_r)$) independent data-streams are used 
in signaling. That is, the $M$-dimensional input 
vector $\bs$ is precoded into an $N_t$-dimensional vector via the 
$N_t \times M$ precoding matrix ${\bf F}$ and transmitted over the channel. 
The discrete-time baseband signal model used is 
\begin{align} 
\label{eq_system_equation} 
\by = {\bf H}  \hsppp {\bf F} \hsppp \bs  + {\bf n} 
\end{align} 
where 
$\by$ is the $N_r$-dimensional received vector, ${\bf H}$ is the $N_{r} \times N_{t}$ 
channel matrix, and ${\bf n}$ is the $N_r$-dimensional zero mean, unit variance 
additive white Gaussian noise. 

\subsection{Channel Model} 
We assume a block fading, narrowband model for the correlation of the channel 
in time and frequency. The main emphasis in this work is on channel correlation 
in the spatial (antennas) domain. The spatial statistics of $\bH$ depend on the 
operating frequency, physical propagation environment which controls the angular 
spreading function and the path distribution, antenna geometry (arrangement and 
spacing) {\emph{etc.}} It is well-known that Rayleigh fading (zero mean complex 
Gaussian) is an accurate model for $\bH$ in a non line-of-sight setting, and hence 
the complete spatial statistics are described by the second-order moments. 

The most general, mathematically tractable spatial correlation model 
is a {\emph{canonical decomposition}}\footnote{This model is referred to as the 
``eigenbeam$/$beamspace model'' in~\cite{canonical_bonek} and is used in 
capacity analysis in~\cite{tulino_ind}.} of the channel along the transmit and 
the receive covariance bases~\cite{canonical_jayesh,canonical_bonek,tulino_ind}. 
In the canonical model, we assume that the auto- and the cross-correlation matrices 
on both the transmitter and the receiver sides have the 
same eigen-bases, and 
therefore we can decompose ${\bf H}$ as 
\begin{eqnarray}
\label{canl}
\bH = \bU_r \hsppp \bH_{  \ind } \hsppp  \bU_t^{\sl H}
\end{eqnarray} 
where $\bH_{\ind}$ has independent, but not necessarily identically 
distributed entries, and $\bU_t$ and $\bU_r$ are unitary matrices. The transmit 
and the receive covariance matrices are given by 
\begin{eqnarray}
\bSigma_t = \bEe \left[ \bH^H \bH \right] & = & 
\bU_t \hsppp \bEe \left[ \bH_{\ind}^H  \bH_{\ind}\right] \hsppp \bU_t^{\sl H} 
= \bU_t \bfLambda_t \bU_t^{\sl H} \nonumber \\
\bSigma_r = \bEe \left[ \bH \bH^H \right] & = & 
\bU_r \hsppp \bEe \left[ \bH_{\ind}  \bH_{\ind}^H \right] \hsppp \bU_r^{\sl H} 
= \bU_r \hsppp \bfLambda_r \hsppp \bU_r^{\sl H} 
\end{eqnarray} 
where $\bfLambda_t = \bEe \left[ \bH_{\ind}^H  \bH_{\ind} 
\right]$ and $\bfLambda_r = \bEe \left[ \bH_{\ind}   \bH_{\ind}^H \right]$ 
are diagonal. Under certain special cases, the model in~(\ref{canl}) reduces to 
some well-known spatial correlation models~\cite{canonical_jayesh}: 
\begin{itemize}
\item 
The case of {\emph{ideal channel modeling}} assumes that the entries of $\bH_{\ind}$ 
are i.i.d.\ standard complex Gaussian random variables~\cite{telatar}. 
The i.i.d.\ model corresponds to an extreme where the channel is characterized 
by a single independent parameter, the common variance. 

\item 
When $\bH_{  \ind }$ is assumed to have the form 
$ 
\frac{1}{\sqrt{ \rho_c } } \cdot 
\bfLambda_r^{1/2} \hsppp \bH_{\iid} \hsppp \bfLambda_t^{1/2}$ with $\bH_{\iid}$ an 
i.i.d.\ channel matrix and the channel power $\rho_c = \trace(\bfLambda_t) = 
\trace(\bfLambda_r)$, 
the canonical model reduces to the often-studied 
normalized 
{\emph{separable correlation framework}} where the correlation of channel entries 
is in the form of a Kronecker product of the transmit and the receive covariance 
matrices~\cite{chuah}. 
The separable model is described by no more than $N_t + N_r$ independent 
parameters corresponding to the eigenvalues $\{ \bfLambda_t(i) \}$ and 
$\{ \bfLambda_r(i) \}$. 

\item 
When uniform linear arrays (ULAs) of antennas are used at the transmitter and the 
receiver, $\bU_t$ and $\bU_r$ are well-approximated by discrete Fourier transform 
(DFT) matrices and the canonical model reduces to the virtual representation
framework~\cite{akbar,akbar_and_venu,venu_capacity}. In contrast to the general 
model in~(\ref{canl}), the virtual representation offers many attractive properties: 
a) The matrices $\bU_t$ and $\bU_r$ are {\em fixed} and independent of the 
underlying scattering environment and the spatial eigenfunctions are beams in the 
virtual directions. Thus, the virtual representation is physically more intuitive 
than the general model in~(\ref{canl}), 
b) It is only necessary that the entries of $\bH_{\ind}$ be 
independent, but not necessarily Gaussian, a criterion important as antenna 
dimensions increase, and c) The case of specular (or line-of-sight) scattering can 
be easily incorporated with the virtual representation framework~\cite{venu_capacity}. 
In contrast to the 
separable model, the virtual representation can support up to $N_t N_r$ independent 
parameters corresponding to the variances of $\{ \bH_{\ind}(i,j) \}$. 
\end{itemize}

While performance analysis is tractable in the i.i.d.\ case, it is 
unrealistic for applications where large antenna spacings or a rich scattering 
environment are not possible. Even though the separable model may be an accurate 
fit under certain channel conditions~\cite{kron_int2}, 
deficiencies acquired by the separability property 
result in misleading estimates of system 
performance~\cite{deficiency_kron1,deficiency_kron2,canonical_jayesh}. The 
readers are referred 
to~\cite{deficiency_kron1,canonical_bonek,zhou} for more details on how the 
canonical$/$virtual models fit measured data better. 

\subsection{Channel State Information} 
If the fading is sufficiently slow, perfect CSI at the receiver is a reasonable 
assumption for practical communication architectures that use a ``training followed 
by signaling'' model. Even in scenarios where this may not be true ({\emph{e.g.,}} 
a highly mobile setting), the performance with imperfect CSI at the receiver can be 
approximated reasonably accurately by the perfect CSI case along with an $\snr$-offset 
corresponding to channel estimation. Thus in this work, we will assume a perfect CSI 
(coherent) receiver architecture. However, obtaining perfect CSI at the transmitter 
is usually difficult due to the high cost associated with channel feedback$/$reverse-link 
training\footnote{In case of Time-Division Duplexed (TDD) systems, the reciprocity 
of the forward and the reverse links can be exploited to train the channel on the 
reverse link. In case of Frequency-Division Duplexed (FDD) 
systems, the channel information acquired at the receiver has to be fed back.}. 

On the other hand, the statistics of the fading process change over much longer 
time-scales and can be learned reliably at both the ends. In addition, recent 
technological advances 
have enabled the possibility of a few bits of quantized channel information to be 
fed back from the receiver to the transmitter at regular intervals. The most common 
form of quantized channel information is via a limited feedback 
codebook ${\cal C}$ of $2^B$ codewords known at both the ends. In this setup, the 
receiver estimates the channel at the start of a coherence block and computes the 
index of the optimal codeword from the codebook ${\cal C}$ for that 
realization of the channel according to some optimality criterion. It then feeds 
back the index of the optimal codeword with $B$ bits over the limited feedback link 
which is assumed to have negligible delay and essentially no errors 
(since $B$ is usually small). The transmitter exploits this 
information to convey useful data over the remaining symbols in the coherence block. 

\subsection{Transceiver Architecture} 
\label{sec_highlight}
The transmitted vector ${\bf F}{\bf s}$ (see~(\ref{eq_system_equation})) 
has a power constraint $\rho$. Assuming that the input symbols ${\bf s}(k)$ have 
equal energy $\frac{\rho}{M}$, the precoder matrix satisfies 
$\trace({\bf F}^H {\bf F}) \leq M$. Non-linear maximum likelihood (ML) 
decoding of the transmitted data symbols using knowledge of ${\bf H}$ at the 
receiver is optimal. However, ML decoding suffers from exponential complexity, in 
both antenna dimensions and coherence length. Thus in practice, a simple linear receiver 
architecture 
like the $\MMSE$ receiver is preferred. 
With this receiver, the symbol corresponding to the $k$-th data-stream 
is recovered by projecting the received signal ${\bf y}$ on to the 
$N_r \times 1$ vector 
\begin{eqnarray}
{\bf g}_k & = & \sqrt{ \frac{\rho}{M} }  \left( \frac{\rho}{M} 
\bH {\bf F} {\bf F}^H {\bf H}^H + {\bf I}_{N_r}  \right)^{-1} 
{\bf H} {\bf f}_{k} 
\end{eqnarray} 
where ${\bf f}_{k}$ is the $k$-th column of ${\bf F}$. That is, the recovered symbol is 
$\widehat{{\bf s}}(k) = \sqrt{ \frac{\rho}{M}} \hsppp {\bf g}_k^H {\bf y}$. The 
signal-to-interference-noise 
ratio ($\sinr$) at the output of the linear filter ${\bf g}_k$ is 
\begin{eqnarray}
\sinr_k =  \frac{ \frac{\rho}{M} |{\bf g}_k^H {\bf H} {\bf f}_{k}|^2  }
{  {\bf g}_k^H \left( \frac{\rho}{M} 
\sum_{i \neq k} {\bf H} {\bf f}_{i} {\bf f}_{i}^H {\bf H}^H + 
{\bf I}_{N_r} \right) {\bf g}_k } 
= \frac{1} { \left( {\bI}_M + 
\frac{ \rho  } { M } {\bf F}^{\sl H} {\bf H}^{\sl H} 
{\bf H} {\bf F} \right)^{-1}_{k,k} } - 1  
\end{eqnarray} 
where the second equality follows from the Matrix Inversion Lemma. 

The outputs $\widehat{{\bf s}}(k)$ 
are passed to the decoder and we assume separate encoders$/$decoders 
for each data-stream, as well as independent interleavers and de-interleavers, 
which reduces the correlation among the interference terms at the outputs of the 
receiver filters. The performance measure is the mutual information between 
${\bf s}$ and $\widehat{{\bf s}}$. Assuming that the interference plus noise 
at the output of the linear filter has a Gaussian distribution, which is true with 
Gaussian inputs and is a good approximation in the non-Gaussian setting when 
$\{ M, N_t, N_r \}$ are large, the mutual information is given by 
\begin{eqnarray}
I({\bf s}; \widehat{ {\bf s} })  =  \sum_{k=1}^M  \log_2 \left( 1 + \sinr_k \right) 
= - \sum_{k=1}^M \log_2 \left( \left( {\bI}_M + 
\frac{ \rho  } { M } {\bf F}^{\sl H} {\bf H}^{\sl H} 
{\bf H} {\bf F} \right)^{-1}_{k,k} \right). 
\end{eqnarray}
When perfect CSI is available at the transmitter and no constraints are 
imposed on the structure of the precoder, the optimal precoder ${\bf F}_{\perf}$ 
is channel diagonalizing 
and is of the form ${\bf F}_{\perf} = \widetilde{ {\bf V}}_{ {\bf H} } 
{\bf \Lambda}_{\sf wf}^{1/2}$ 
where ${\bf V}_{\bH} {\bf \Lambda}_{\bH} {\bf V}_{\bH}^H$ is an eigen-decomposition of 
$\bH^H \bH$ with the eigenvalues arranged in non-increasing order, 
$\widetilde{ {\bf V} } _{\bH}$ is the $N_t \times M$ principal submatrix of 
${\bf V}_{\bH}$, and $\bfLambda_{ {\sf wf} }$ is an $M \times M$ matrix with 
non-negative entries only along the leading diagonal and these entries are obtained 
by waterfilling. In this setting, the mutual information is given by 
\begin{eqnarray}
I_{\perf}({\bf s}; \widehat{ {\bf s} }) = \sum_{k = 1}^M \log_2 \left( 1 + 
\frac{\rho}{M} {\bf \Lambda}_{\bH}(k) {\bf \Lambda}_{\sf wf}(k)  \right).
\end{eqnarray}
The optimality of ${\bf F}_{\perf}$ with other choices of objective functions is 
also known; see 
\cite{lee_petersen,salz,yang_roy,scaglione_gia_barb,sampath,sampath2,yang_roy2,scaglione_spbgs,palomar_precode}. 

\subsection{Limited Feedback Framework}
The focus of this work is on understanding the implications of partial CSI at the 
transmitter on the performance of the precoding scheme. In particular, there exists a 
codebook of the form ${\cal C} = \{  {\bf F}_i, i = 1, \cdots , 2^B \}$ where 
${\bf F}_i$ is an $N_t \times M$ precoder matrix with $\trace({\bf F}_i^H {\bf F}_i) 
\leq M$. The most general structure for 
${\bf F}_i$ is ${\bf F}_i = {\bf V}_i {\bf \Lambda}_i^{1/2}$ where ${\bf V}_i$ is an 
$N_t \times M$ semiunitary matrix and ${\bf \Lambda}_i$ is an $M \times M$ non-negative 
definite, diagonal power allocation matrix. While the structure of the 
optimal limited feedback codebook of $B$ bits could involve allocating some fraction 
of $B$ to the power allocation component of ${\bf F}_i$, numerical studies indicate 
that the degradation in performance is minimal when ${\bf \Lambda}_i$ is chosen to 
be fixed (say, ${\bf \Lambda}_{\stat}$ with $\trace({\bf \Lambda}_{\stat}) \leq M$), 
but designed appropriately, as a function of $\snr$ if necessary, so that it can 
be easily adapted to statistical variations without recourse to Monte Carlo 
methods\footnote{The design of ${\bf \Lambda}_{\stat}$ will be dealt with in 
Sec.~\ref{sec4}.}. 

Motivated by this heuristic, in this work, all the $B$ bits in limited feedback are 
allocated to quantize the eigenspace of the channel. That is, the codebook is 
${\cal C} = \{ {\bf V}_i : {\bf V}_i^H {\bf V}_i = {\bf I}_M \}$ and the index of the 
codeword that is fed back is 
\begin{eqnarray}
j^{\star} = \arg \max_j \left\{  
- \sum_{k=1}^M \log_2 \left( \left( \bI_M + \frac{\rho}{M} 
{\bf \Lambda}_{\stat}^{1/2} {\bf V}_i^H {\bf V}_{\bH} {\bf \Lambda}_{\bH} {\bf V}_{\bH}^H 
{\bf V}_i {\bf \Lambda}_{\stat}^{1/2} \right)^{-1}_{k,k}  \right) \right\}.  
\end{eqnarray} 
Although computing $j^{\star}$ is straightforward, the design of an optimal codebook to 
maximize $I({\bf s}; \widehat{{\bf s}})$ seems difficult. Here, we adopt a suboptimal 
strategy where the goal is to maximize the average projection of the best codeword from 
${\cal C}$ onto $\widetilde{{\bf V}}_{\bH}$. Towards the precise mathematical 
formulation of this problem, we need a metric to define distance between two semiunitary 
matrices. 
 
\subsection{Distance Metrics and Spherical Caps on the Grassmann Manifold} 
We now 
recall some well-known facts about the Grassmann manifold. The unit sphere 
in ${\mathbb C}^{N_{t}}$, also known as the uni-dimensional\footnote{Uni-dimensional 
because its definition is based on the norm of an $N_t \times 1$ vector.} 
complex Stiefel manifold $\St(N_t,1)$, is defined as $\St(N_t,1) = 
\left\{ {\bf x} \in {\mathbb C}^{N_{t}} : \|{\bf x}\| = 1\right\}$. The invariance of 
any vector ${\bf x}$ to transformations of the form ${\bf x} \mapsto e^{j\phi}{\bf x}$ 
in the above definition is incorporated by considering vectors modulo the above map. 
The partitioning of $\St(N_t,1)$ by this equivalence map results in the 
uni-dimensional Grassmann manifold ${\cal{G}}(N_t,1)$. 
In short, the Grassmann manifold corresponds to a linear subspace 
in an Euclidean space. Similarly, the class of $N_t \times M$ 
semiunitary matrices forms the $M$-dimensional complex Stiefel manifold $\St(N_t,M)$ 
and points on the $M$-dimensional complex Grassmann manifold ${\cal{G}}(N_t,M)$ 
are identified modulo the $M$-dimensional unitary space. 

A literature survey of packings on ${\cal{G}}(N_t,1)$~\cite{barg,arias_smith,sloane} 
shows that many distance metrics are equivalent to the dot product metric which is 
the most natural metric 
from an engineering perspective. The dot product metric is defined as 
$d\left({\bf x}_{1},{\bf x}_{2}\right) = \sqrt{1-|{\bf x}_{1}^{\sl H}{\bf x}_{2}|^2}$. 
Using this distance metric, for any $\gamma < 1$, we can define a {\emph{spherical 
cap}} with center ${\bf o}$ and radius $\gamma$ (as a {\em submanifold} 
on ${\cal{G}}(N_t,1)$) 
as the open set ${\mathbb O}({\bf o},\gamma) = \left\{{\bf x} \in {\cal{G}}(N_t,1) : 
d\left({\bf x},{\bf o}\right) < \gamma \right\}.$ A spherical cap on ${\cal{G}}(N_t,1)$ 
induces a spherical cap on $\St(N_t,1)$ via the equivalence partitioning 
generated by the map ${\bf x} \mapsto e^{j\phi}{\bf x}$. 

In the more general $M > 1$ case, there is no unique distance metric extension. 
While various well-defined distance metrics can be pursued, we will focus on the 
{\emph{projection $2$-norm distance metric}}~\cite{arias_smith}. 
Here, the distance between two $N_t \times M$ semiunitary matrices $\bFa_1$ and 
$\bFa_2$ is defined as 
\begin{eqnarray}
d_{\proj, \hsppp 2}(\bFa_1, \bFa_2) = \lambda_{\max} \left( 
\bFa_1 \bFa_1^H - \bFa_2 \bFa_2^H \right). 
\end{eqnarray} 
A particular choice of the distance metric is not extremely critical in precoder 
optimization since codebooks designed with different choices of distance metrics 
result in near-identical performance~\cite{david_heath_multimode,stat_semiunitary}. 
In addition to this fact, the following 
lemma shows that the projection $2$-norm metric is attractive by being a natural 
generalization of the dot product metric. 
\begin{lem} 
\label{lem_dist}
In the $M = 1$ case, the projection $2$-norm metric reduces to the standard 
dot product metric. 
\end{lem}
\begin{proof}
Let ${\bf v}_1$ and ${\bf v}_2$ be two unit-normed $N_t \times 1$ vectors. 
Then, the projection $2$-norm distance between ${\bf v}_1$ and ${\bf v}_2$ 
is defined as $d_{\proj,\hsppp 2}({\bf v}_1, {\bf v}_2) = \lambda_{\max} \left( 
{\bf v}_1 {\bf v}_1^H - {\bf v}_2 {\bf v}_2^H \right)$. 
We can write the matrix within the $\lambda_{\max}(\cdot)$ operation as 
$\left[ {\bf v}_1 {\hspace{0.06in}} {\bf v}_2 \right] 
\left[ {\bf v}_1 {\hspace{0.06in}} -{\bf v}_2 \right]^H$. 
Since the non-trivial eigenvalues of a matrix product ${\bf AB}$ are the same as those 
of ${\bf BA}$, we need the largest eigenvalue of 
\begin{eqnarray}
{\bf X} =  \left[ \begin{array}{l} 
{\bf v}_1^H \\ -{\bf v}_2^H 
\end{array} 
\right]
\left[ \begin{array}{cc}
{\bf v}_1 & {\bf v}_2 \end{array}
\right] = 
\left[ \begin{array}{cc} 
1 & {\bf v}_1^H {\bf v}_2 \\ 
-{\bf v}_2^H {\bf v}_1 & -1 
\end{array}
\right]. 
\end{eqnarray}
Expanding the characteristic equation of ${\bf X}$, $\det({\bf X} - 
\lambda {\bf I}_2 ) = 0$, we have $\lambda^2 = 1 - |{\bf v}_1^H {\bf v}_2|^2$. 
Using the positive root for $\lambda_{\max}$, the lemma follows immediately. 
\end{proof}
\begin{prop} 
\label{prop_000}
We now state some properties of the projection $2$-norm metric: 
\newline 
\noindent 
1) $0 \leq d_{\proj, \hsppp 2}(\bFa_1, \bFa_2) \leq 1$, 
\newline 
\noindent 
2) More precisely, 
$d_{\proj, \hsppp 2} (\bFa_1, \bFa_2) = 
\sqrt{1 - \lambda_{\min}( \bFa_1^H \bFa_2 \bFa_2^H \bFa_1  )},$ 
and 
\newline 
\noindent 
3) Equality in the lower bound of 1) occurs if and only if $\bFa_1 = \bFa_2$ on 
${\cal G}(N_t, M)$ while equality is possible in the upper bound if and only if 
$\lambda_{\min}( \bFa_1^H \bFa_2 \bFa_2^H \bFa_1  ) = 0$. 
\end{prop} 
\begin{proof} 
The proof is provided in three parts. 
\newline 
\noindent 1) 
Using the fact that $\bFa_1 \bFa_1^H - \bFa_2 \bFa_2^H$ is Hermitian and its trace 
equals zero, we see that $\lambda_{\max}( \bFa_1 \bFa_1^H - \bFa_2 \bFa_2^H ) < 0$ 
is impossible. For the upper bound, note that 
\begin{eqnarray}
\lambda_{\max}( \bFa_1 \bFa_1^H - \bFa_2 \bFa_2^H ) \leq 
\lambda_{\max}(\bFa_1 \bFa_1^H ) = 
\lambda_{\max}(\bFa_1^H \bFa_1) = 1.
\end{eqnarray}
\noindent 2) We the need the following result~\cite{silvester} that helps in 
computing the determinant of partitioned matrices. 
\begin{lem} 
\label{lem_partition}
If ${\bf X}, {\bf Y}, {\bf Z}$ and ${\bf W}$ are $n \times n$ matrices and ${\bf W}$ 
is invertible, we have 
\begin{eqnarray} 
\det \left[ \begin{array}{cc}
{\bf X} & {\bf Y} \\ 
{\bf Z} & {\bf W} 
\end{array}  \right] = \det({\bf X} - {\bf Y} {\bf W}^{-1} {\bf Z}) 
\cdot \det({\bf W}). 
\end{eqnarray}
\endproof 
\end{lem} 
Using the above fact and the trick (in Lemma~\ref{lem_dist}) of rewriting the 
eigenvalues of ${\bf A}{\bf B}$ in terms of eigenvalues of ${\bf B}{\bf A}$, 2) 
follows trivially. 
\newline 
\noindent 3) If $d_{\proj, \hsppp 2}(\bFa_1, \bFa_2) = 0$, then it is easy to see 
that $\bFa_1 \bFa_1^H = \bFa_2 \bFa_2^H$ from which we note that 
$\bFa_1 = \bFa_2 \bFa_2^H \bFa_1$. Observe that $\bFa_2^H \bFa_1$ is $M \times M$ 
and unitary, and hence, $\bFa_1 = \bFa_2$ on ${\cal G}(N_t, M)$. The other direction 
of the statement follows trivially. Both the directions of the upper bound 
follow from the expression in 2). 
\end{proof}

The trick in proving Lemma~\ref{lem_dist} and statement 2) in Prop.~\ref{prop_000} 
is useful and will be used again 
in the construction of the scaling map (see Appendix~\ref{app_scale}). 
Once a choice of distance metric has been settled, the definition of a spherical cap 
with center ${\bf O}$ and radius $\gamma$ (as a submanifold on ${\cal{G}}(N_t,M)$) 
follows naturally as the open set ${\mathbb O}({\bf O},\gamma) = 
\left\{{\bf X} \in {\cal{G}}(N_t,M) : d_{\proj, \hsppp 2}\left({\bf X},{\bf O}\right) 
< \gamma \right\}.$ The codebook design problem can now be simply stated as: 
\begin{eqnarray}
{\rm Construct} {\hspace{0.05in}} {\cal C} = \{ {\bf V}_i, i = 1, \cdots, 2^B \} 
{\hspace{0.05in}} {\rm s.t.} \hspp
\bEe_{\bH} \left[ \min_{i=1 , \hsppp \cdots , \hsppp 2^B} \hspp 
d_{\proj, \hsppp 2}({\bf V}_i, \widetilde{ {\bf V} }_{\bH}) \right]  
{\hspace{0.05in}} {\rm is} {\hspace{0.05in}} {\rm minimized}. \nonumber 
\end{eqnarray}
We now work towards a systematic codebook construction for this problem.

\ignore{ 
\begin{prop} 
\label{prop_opt_precoder}
Let $\mse = [ \mse_1 \hspp \cdots \hspp \mse_M]$ with $\mse_i$ denoting the mean 
squared error of the $i$-th data-stream. Then, the choice of ${\bf F}_{\opt}$ 
that 1) minimizes $f(\mse)$ for any Schur-concave$/$convex function $f(\cdot)$ 
that is also monotonically increasing in its arguments, 2) maximizes the mutual 
information, or 3) minimizes $\sum_{k=1}^M h (\mse_k)$ for any $h(\cdot)$ that is a 
continuous, increasing, and convex function of its argument is of the form 
\begin{eqnarray} 
\label{fopt1}
{\bf F}_{\opt} = \exp(j \theta) \hsppp \left[ 
\bv_1 \hsppp \cdots \hsppp \bv_M \right] \hsppp \bGamma  
\end{eqnarray} 
where $\bGamma$ is an appropriately chosen unitary matrix, $\bv_i$ is the 
$i$-th column of ${\bf V}_{\bf H}$, and $\theta \in {\mathbb{R}}$ is 
some fixed number. Here, we assume an SVD of ${\bf H}$ to be ${\bf H} = 
{\bf U}_{\bf H} {\bf \Lambda}_{\bf H} {\bf V}_{\bf H}^{\sl H}$ with the 
non-trivial singular values ${\bf \Lambda}_{\bf H}(i)$ arranged in decreasing 
order. 
\end{prop} 
}

\section{Matched Versus Mismatched Channels} 
\label{sec3} 
The case of unstructured precoding with genie-aided perfect CSI was summarized 
in Sec.~\ref{sec_highlight} which resulted in ${\bf F}_{\perf} = 
\widetilde{ {\bf V}}_{ {\bf H} } {\bf \Lambda}_{\sf wf}^{1/2}$. The construction 
of $\widetilde{ {\bf V}}_{ {\bf H} }$, as well as ${\bf \Lambda}_{\sf wf}$, 
necessitates the tracking of the channel evolution which is difficult. To avoid 
this problem and to reduce the complexity of precoding, the following 
structured precoding was introduced in~\cite{stat_semiunitary}. 

\begin{itemize} 
\item 
When the precoder is assumed to be structured as ${\bf F} = {\bf V} 
{\bf \Lambda}_{\stat}^{1/2}$ with ${\bf V}$ an $N_t \times M$ semiunitary matrix, 
and ${\bf \Lambda}_{\stat}$ an $M \times M$ fixed, ${\sf rank}$-$M$ power allocation 
matrix, 
the optimal choice of ${\bf V}$ under perfect CSI is $\widetilde{{\bf V}}_{\bH}$. 
This optimality is assured for many different classes of objective functions 
apart from the case of maximizing 
mutual information. When only statistical information is available 
at the transmitter, the optimal choice of ${\bf V}$ is ${\bf V}_{\stat}$ where 
${\bf V}_{\stat}$ is a set of $M$ dominant eigenvectors of ${\bf \Sigma}_t$, 
the transmit covariance matrix. We call these two schemes {\em optimal and 
statistical structured precoding schemes}, respectively.  

\item 
We study the performance loss between these two schemes as a function of the 
channel statistics. When one antenna dimension grows to infinity at a rate faster 
than the other\footnote{That is, when $\frac{N_t}{N_r} \rightarrow 0$ or $\infty$ 
as $\{ N_t, N_r\} \rightarrow \infty$.}, which we refer to as the {\em relative antenna 
asymptotics} case, channel hardening 
leads to convergence of the right singular values of the channel to the 
eigenvalues of ${\bf \Sigma}_t$ and hence, ensures that the statistical scheme 
performs near-optimally. This conclusion generalizes prior results in the 
beamforming case where statistical beamforming is shown to be near-optimal in 
the relative antenna asymptotics setting~\cite{raghavan_spl_issue}.

\item 
Further, for any reasonably large (but fixed) value of antenna dimensions, the 
relative performance loss between the two schemes is minimized by the following 
choice of statistics: 1) The set of transmit eigenvalues $\{ \bfLambda_t(i) \}$ can 
be partitioned into two components: a well-conditioned component of $M$ dominant 
eigenvalues, and the remaining $N_t - M$ transmit eigenvalues are ill-conditioned 
away from the dominant set, and 2) The set of receive eigenvalues $\{ \bfLambda_r(i) \}$ 
are well-conditioned. In particular, if $\trace(\bSigma_t) = \trace(\bSigma_r) = 
N_t N_r$, the structure of $\bfLambda_t$ and $\bfLambda_r$ that 
minimizes performance loss is $\bfLambda_t(1) = \cdots = \bfLambda_t(M) = 
\frac{N_t N_r}{M}, \bfLambda_t(M+1) = \cdots = \bfLambda_t(N_t) = 0$, and 
$\bfLambda_r(1) = \cdots = \bfLambda_r(N_r) = N_t$. Such a channel is said to 
be {\emph{matched}} to the precoding scheme. On the other extreme, 
statistical structured precoding in an i.i.d.\ channel 
leads to very high performance loss when compared with the optimal scheme. 
Thus, an i.i.d.\ channel is {\emph{mismatched}} to the precoding scheme. 
More important to note is that any feedback (limited or 
otherwise) is helpful {\em only} in mismatched channels and only when 
the transmit and the receive dimensions are proportionate. 
This conclusion is a generalization of our earlier beamforming 
result~\cite{raghavan_spl_issue}.

\end{itemize} 
The readers are referred to~\cite{stat_semiunitary} for details. Henceforth, the 
focus will be on mismatched channels primarily because the potential to bridge the 
performance gap between the statistical and perfect CSI schemes is maximum. Our 
goal is to construct a systematic, statistics-dependent codebook (of a 
fixed size $2^B$) that ensures this bridging.

\section{Quantized Feedback Designs to Bridge the Performance Gap}
\label{sec4} 
In contrast to the i.i.d.\ case where the isotropicity of the right singular 
subspace of the channel leads to a design~\cite{david_heath_multimode} based on 
Grassmannian subspace packings~\cite{sloane}, spatial correlation skews this 
isotropicity and poses fundamental challenges. The study of statistical precoding 
motivates the following heuristic in the correlated case. While the 
asymptotic channel hardening (and the consequent near-optimality of statistical 
precoding) does not carry over when $N_t$ and $N_r$ are small or when they are 
proportionate, it is expected that the distance between ${\bf V}_{\stat}$ and 
$\widetilde{\bf V}_{\bH}$ is small on average. Thus, when we have the freedom to 
pick more than one codeword ($B > 0$), the codewords should correspond to a ``local 
quantization'' of ${\bf V}_{\stat}$. The notion of local quantization will be made 
precise shortly. 

We now describe the codebook design for limited feedback precoding. Our design is a 
multi-mode generalization of the beamforming codebook proposed 
in~\cite{raghavan_spl_issue,vasanth_isit06}. The differences between the two schemes 
lie in packing subspaces, rather than lines, and in the choice of an appropriate 
distance metric. 
For this, we introduce the notion of {\emph{generalized eigenvalues}} of subspaces of 
${\bf \Sigma}_t$. Consider the family of subspaces spanned by $M$ distinct 
eigenvectors of ${\bf \Sigma}_t$. Note that there are ${N_t \choose M}$ members 
in this family. For each such subspace, we associate a {\emph{generalized 
eigenvalue}} defined as the $M$-fold product of the corresponding transmit 
eigenvalues. For example, if $N_t = 4$ and $M = 2$ with the columns of 
${\bf U}_t$ denoted by ${\bf u}_i, i = 1, \cdots , 4$, the six subspaces 
correspond to the $4 \times 2$ matrices: $[\bu_1 \hsppp \bu_2], [\bu_1 \hsppp \bu_3], 
[\bu_1 \hsppp \bu_4], [\bu_2 \hsppp \bu_3], [\bu_2 \hsppp \bu_4]$ and 
$[\bu_3 \hsppp \bu_4]$. The generalized eigenvalue corresponding to 
$[\bu_1 \hsppp \bu_2]$ is $\bfLambda_t(1) \bfLambda_t(2)$ {\em etc}. 
Note that among all the $M$-dimensional subspaces of ${\bf \Sigma}_t$, the subspace 
spanned by ${\bf V}_{\stat}$ results in the largest generalized eigenvalue.

The proposed codebook design has three components: 1) a statistical component, 
2) local perturbation components, and 3) an RVQ component. The cardinalities of 
these components are denoted by $N_{\stat}, N_{\loc}$ and $N_{\glo}$ with the 
feedback rate defined by $B = \log_2( N_{\stat} + N_{\loc} + N_{\glo} )$. 

\noindent{\bf{\emph{Statistical Component:}}} While the distance between 
${\bf V}_{\stat}$ and ${\widetilde{\bf V}}_{\bH}$, an instantaneous realization of the 
$M$-dominant right singular vectors of the channel is expected to be 
small on average, the 
precise probability distribution of this distance is determined by the separation 
(gap) between the generalized eigenvalues of ${\bf \Sigma}_t$. For example, 
if the first two dominant generalized eigenvalues are close to each other, there 
is a non-negligible probability for the event that the best quantizer is the 
subspace whose generalized eigenvalue is the smaller of the two and hence, the 
distance between ${\bf V}_{\stat}$ and the optimal precoder 
could be arbitrarily close\footnote{Note from Prop.~\ref{prop_000} that the distance 
between the first two dominant eigen-spaces of ${\bf \Sigma}_t$ is $1$. This is 
because $\lambda_{\min}({\bf V}_1^H {\bf V}_2 {\bf V}_2^H {\bf V}_1) = 0$ 
where ${\bf V}_1$ and ${\bf V}_2$ 
denote the first two dominant eigen-spaces.} to $1$. On the other hand, if the 
largest generalized eigenvalue of ${\bf \Sigma}_t$ is much larger than the other 
generalized eigenvalues, the probability distribution of this distance is concentrated 
around zero. Thus the gap between the largest generalized eigenvalue and the 
other generalized eigenvalues heuristically determines 
the cardinality of the statistical component, $N_{\stat}$. In our design, 
a threshold $\beta$ is chosen {\emph{a priori}} for the generalized 
eigenvalues and the statistical component consists of all $M$-dimensional subspaces 
such that their generalized eigenvalue exceeds the threshold. That is, 
the {\emph{statistical component}} is the set ${\cal S} = 
\left\{ i : \frac{\mu_i}{\mu_1} > \beta \right\}$ where $\mu_i$ are the $M$-fold 
generalized eigenvalues of ${\bf \Sigma}_t$ and $\mu_1$ is the largest generalized 
eigenvalue. The cardinality of ${\cal S}$ is $N_{\stat}$. 

\begin{figure}[htb!]
\centering
\begin{tabular}{c}
\includegraphics[height=3.2in,width=4in]{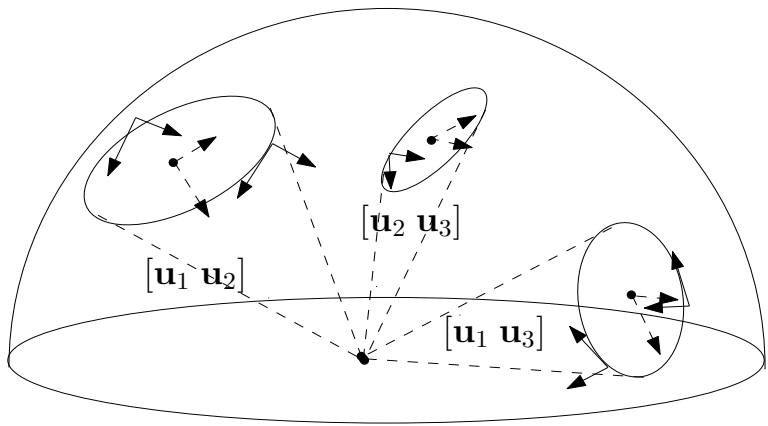}
\end{tabular}
{\vspace{-0.9in}}
\caption{Proposed Codebook Design for $N_t = 3, M = 2$, and $B = 3$ with only the 
statistical and local components.}\label{fig_cbo}
\end{figure}

\noindent{\bf{\emph{Local Components:}}} For the $i$-th member of the statistical 
component, we construct $N_{\loc}^i$ codewords so that they are localized and 
well-packed around the corresponding statistical codeword. While these local 
codewords can theoretically be designed via VQ, we provide low-complexity 
alternatives in Sec.~\ref{sec_sca} where we also elaborate on the notions of 
{\em localized and well-packed}. The choice of $N_{\loc}^i$ is 
in proportion to the generalized eigenvalue of the subspace. The heuristic behind 
this choice is as follows: The larger the separation of the generalized eigenvalue 
$\mu_1$ (corresponding to ${\bf V}_{\stat}$) from the next largest generalized 
eigenvalue or the more matched ${\bf \Sigma}_t$ is, the lesser the relevance of the 
less-dominant subspaces in terms of precoding and hence, the smaller the values of 
$\{ N_{\loc}^i \}, i > 1$ need to be. These $N_{\loc} = 
\sum_{i =1}^{N_{\stat} } N_{\loc}^i$ codewords form the {\emph{local component}} of 
our codebook design. 

In Fig.~\ref{fig_cbo}, we illustrate the design of a codebook with statistical and 
local components where $N_t = 3, M = 2$, $N_{\stat} = 3$, 
$N_{\loc}^1 = N_{\loc}^2 = 2$ and $N_{\loc}^3 = 1$. If ${\bf U}_t = 
[{\bf u}_1 \hsppp {\bf u}_2 \hsppp {\bf u}_3]$, then the three statistical transmit 
eigenspaces with $M = 2$ are those spanned by $[{\bf u}_1 \hsppp {\bf u}_2]$, 
$[{\bf u}_1 \hsppp {\bf u}_3]$ and $[ {\bf u}_2 \hsppp {\bf u}_3]$. The 
``directions'' corresponding to these subspaces are symbolically represented in 
the figure with dashed lines. The first local component consists of two codewords 
around $[{\bf u}_1 \hsppp {\bf u}_2]$ and so on. Since there are eight 
codewords in our design, this codebook can be 
parameterized with $B = 3$ bits.

\noindent{\bf{\emph{RVQ Component:}}} 
If $B$ is sufficiently large, there is a need to refine the quantization of 
$\widetilde{{\bf V}}_{ {\bf H}}$. In this setting, $N_{\glo} \triangleq 
2^B - N_{\stat} - N_{\loc}$ random channel matrices are generated according to 
the relationship in~(\ref{canl}) and their $M$-dominant right singular vectors are 
used as the semiunitary precoder codewords in the RVQ component. 
Note that the RVQ component can be generated with low-complexity once the 
statistics are known perfectly. 

\subsection{Power Allocation} 
It is preferred that the power allocation matrix ${\bf \Lambda}_{\stat}$ be only 
dependent on the channel statistics and be easily adaptable to statistical variations. 
The optimal choice of $\bfLambda_{\stat}$ needs to be constructed via a Monte Carlo 
algorithm which is difficult to implement as well as adapt to statistical variations 
with low-complexity. As an alternative, we consider three low-complexity power 
allocations: 1) uniform power allocation across 
the excited modes, 2) waterfilling based on ${\bf \Lambda}_t(i), 
\hspp i = 1, \hsppp \cdots, \hsppp M$, and 3) power allocation proportional to 
the transmit eigenvalues. The last two schemes have near-identical performances 
and are near-optimal in the low-$\snr$ regime while 
uniform power allocation is more useful in the high-$\snr$ regime. 

\subsection{Codeword Selection}
The receiver acquires the channel information at the start of a coherence block and 
it computes the index of the optimal codeword from the codebook that maximizes the 
instantaneous mutual information. The receiver then communicates to the transmitter 
the index of the optimal codeword with $B$ bits. The transmitter uses the optimal 
codeword along with an appropriate power allocation to communicate over the remaining 
period in the coherence block.

\begin{figure}[htb!]
\begin{center}
\begin{tabular}{c}
\centerline{\includegraphics[width=5.5in,height=2in]{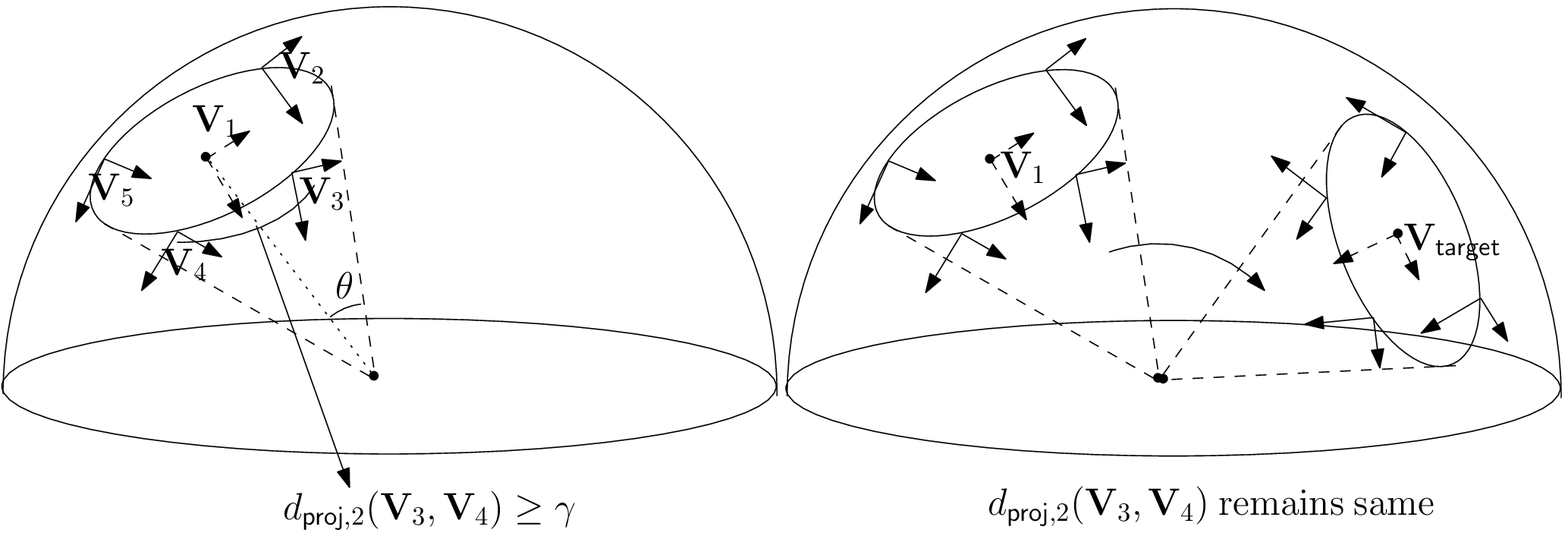}}
\\ (a) 
\\ 
\includegraphics[width=5.5in,height=2in]{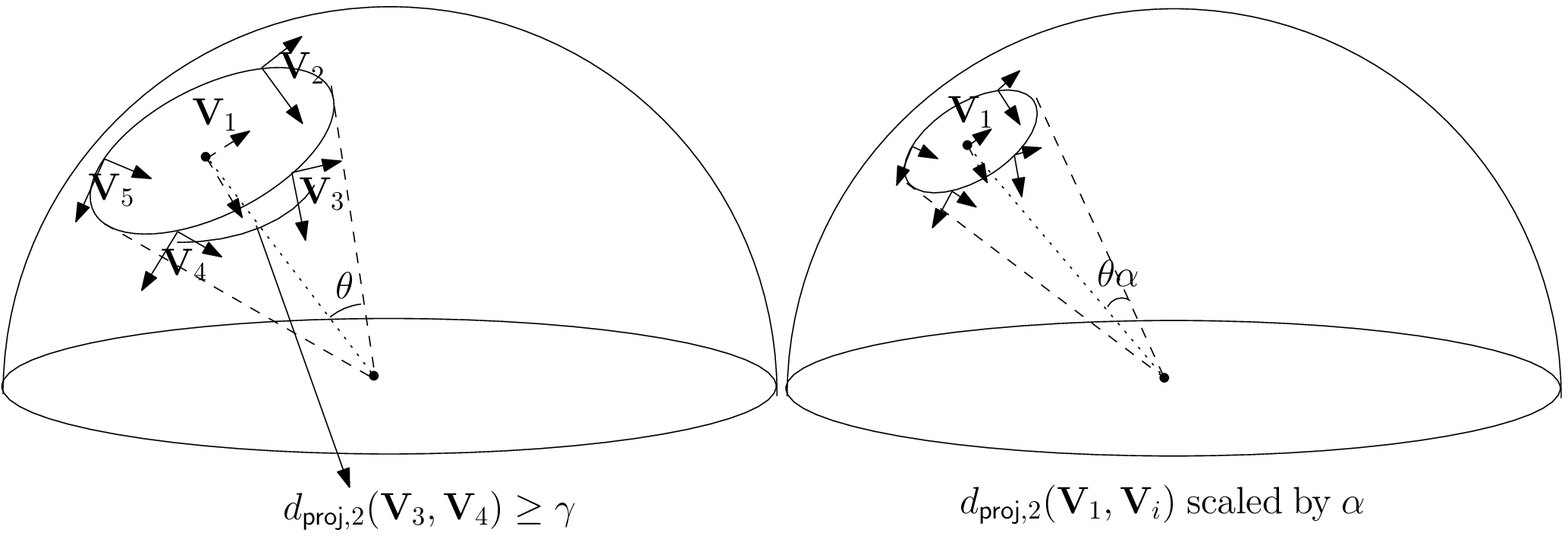}\\
(b) 
\end{tabular}
\caption{\label{fig_maps}(a) Rotation of a root codeset of semiunitary 
precoders $\{ {\bf V}_i, \hsppp i = 1, \cdots, 5 \}$ with $N_t = 3$ and 
$M = 2$. The root codeset satisfies the localization and well-packing properties 
described in Sec.~\ref{sec_sca}. The distance between any two precoders remains 
unchanged after rotation. 
(b) Scaling of the root codeset by $\alpha$. The position of ${\bf V}_1$ remains 
unchanged after scaling.}
\end{center}
\vspace{-5mm}
\end{figure}

\section{Rotating and Scaling Spherical Caps on ${\cal{G}}(N_t,M)$} 
\label{sec_sca} 

We now propose mathematical maps to ensure that the codebook design proposed 
above can be realized with low-complexity. For this, we need the notion of a 
{\em root codeset.} Let ${\cal{R}} = \{ \bFa_i , i = 1, \cdots, N \}$ be a root 
codeset\footnote{We use the term root codeset to indicate that the construction 
of ${\cal C}$ is rooted in the design of a `good' ${\cal R}$.} of $N$ semiunitary 
matrices satisfying the following properties which are characteristic of a `good' 
local quantization: 
\begin{enumerate}
\item 
{\bf{\emph{Localization:}}} The root codeset is localized (centered) around $\bFa_1$. 
That is, there exists a $\theta \in (0,1)$ such that 
$d_{\proj,\hsppp 2}(\bFa_1 , \bFa_i) \leq \theta$ for all $i \neq 1$. The 
smaller the value of $\theta$, the more localized a packing. 
We often label ${\bf V}_1$ as the {\emph{center of 
the root codeset}}. This is illustrated in Fig.~\ref{fig_maps} where a set of 
$N = 5$ precoders form the localized root codeset in the $N_t = 3, M = 2$ setting. 
\item 
{\bf{\emph{Well-Packing:}}} The codewords in ${\cal{R}}$ are well-packed 
(well-separated). That is, given some $\gamma \in (0, \gamma_{\max}(N_t, M, N, \theta))$, 
the minimum distance of the packing 
$d_{\min}({\cal{R}})$ defined as $d_{\min}({\cal{R}}) \triangleq \min_{i \neq j} 
d_{\proj,\hsppp 2}(\bFa_i , \bFa_j)$ is larger than $\gamma$. 
The larger the value of $\gamma$, the well-packed ${\cal R}$ is. Hence 
$\gamma$ can also be viewed as a measure of the packing density. 
Here, $\gamma_{\max}(N_t,M,N, \theta)$ is the maximum possible packing 
density\footnote{While the exact characterization of $\gamma_{\max}(N_t,M,N,\theta)$ 
remains an open problem for general values of $N_t, M$, $N$ and $\theta$, some bounds 
have been established; 
see~\cite{sloane,arias_smith,xia_corr,david_grass} and references therein.} 
achievable in the Grassmann manifold ${\cal{G}}(N_t,M)$ with $N$ codewords 
localized in a cap of radius $\theta$. 
\end{enumerate} 
Note that for any fixed choice of $N_t, M$ and $N$, it is intuitive to expect that 
$\gamma_{\max}(N_t, M, N, \theta)$ decreases as $\theta$ decreases. In other words, 
the above two properties are in some sense conflicting with a root codeset that is 
more localized necessarily forced to have a small packing density and {\em vice 
versa}. 

Despite this apparent difficulty, it is important to note that a packing 
with the above properties can {\em always} be constructed, either via 
algebraic methods or via a vector quantization~\cite{honig,bhaskar_rao} approach 
(that is, a brute force search via Monte Carlo-type algorithms). 
Furthermore, 
${\cal R}$ needs to be constructed (offline) just once, and once this has been done, 
${\cal C}$ can be designed for any statistics starting from ${\cal R}$. For this, we 
now show how mathematical operations can be constructed to perform the following two 
tasks: 
\newline \noindent 
1) Given a root codeset ${\cal{R}}$ of $N$ codewords with a packing density 
$\gamma$ and a target center $\bFa_{\target}$, how can we center ${\cal{R}}$ around 
$\bFa_{\target}$ without having to resort to a VQ-type codebook construction 
{\emph{again}}? That is, we seek a map to rotate the center of ${\cal{R}}$ to 
$\bFa_{\target}$ without changing the packing density, and 
\newline \noindent 
2) Given a root codeset ${\cal{R}}$ centered around $\bFa_1$ with a packing density 
of $\gamma$ and some fixed $\alpha \in (0,1)$, how can we scale ${\cal{R}}$ so that 
the packing density of the resultant codeset is $\alpha \gamma$? That is, we seek a 
map to reduce the minimum distance of ${\cal{R}}$ without changing its center. 

While we develop such maps for spherical caps$/$submanifolds, we will state the 
results as applicable to finite element subsets of ${\cal{G}}(N_t,M)$. But prior 
to that, we recall results from a recent work~\cite{samanta_heath} where rotation 
and scaling maps to solve 1) and 2) (as above) have been proposed 
in the beamforming case ($M = 1$). 
The rotation map is straightforward and is effected by an 
appropriately chosen unitary matrix. In contrast to the rotation operation, the 
scaling map requires some care due to the constraints of the space. For example, 
an operation of the form ${\bf x} \mapsto \alpha{\bf x}$ where $\alpha \in {\mathbb{R}}$ 
yields a vector that is not unit-norm. It is to be noted that both rotation and 
scaling maps are non-unique. We summarize the map 
of~\cite{samanta_heath} in the following lemma\footnote{The readers are 
referred to~\cite{raghavan_spl_issue} for details of the proof.} for $M = 1$. 

\begin{lem}[See~\cite{samanta_heath}] 
\label{lem_scale} 
Let ${\cal{R}} = \{ \bfa_i, i = 1, \cdots, N \}$ be a root codeset in 
${\cal{G}}(N_t,1)$ with a packing density $\gamma$ and center $\bfa_1$. The map 
that effects the rotation of $\bfa_1$ to $\bfa_{\target}$ is given by 
$r({\bfa_i}) = {\bf U}_{\target} \bfa_i$ with ${\bf U}_{\target}$ 
satisfying\footnote{One 
possible choice of ${\bf U}_{\target}$ is ${\bf U}_{\target} = 
\left[{\bf v}_{\target} \hspp {\bf v}_{\target}^{\perp} \right] 
\hsppp \left[ \begin{array}{cc} {\bf v}_1 & {\bf v}_1^{\perp} 
\end{array} \right]^H$ where ${\bf v}_{\target}^{\perp}$ and ${\bf v}_1^{\perp}$ 
refer to matrix representatives from the $N_t \times (N_t-1)$ dimensional null-space 
of ${\bf v}_{\target}$ and ${\bf v}_1$, respectively. That is, 
${\bf v}_1^{\perp, \hsppp H} {\bf v}_1^{\perp} = {\bf I}_{N_t - 1}$ and 
${\bf v}_1^{\perp, \hsppp H} {\bf v}_1 = {\bf 0}_{N_t - 1 \times 1}$.} 
${\bfa}_{\target} = {\bf U}_{\target} \bfa_1$. 
For scaling by $\alpha$, we first define a rotation map $r_{\vertex}$ generated by a 
unitary matrix ${\bf U}_{\vertex}$ that effects the rotation of the center ${\bf v}_1$ 
to ${\bf v}_{\vertex} = [1, \hsppp 0, \cdots, 0]^{\sl T}$, a vertex of the unit cube. 
Then, define a vertex scaling map $s_{\vertex} : 
{\mathbb O}({\bf v}_{\vertex},\gamma) \mapsto {\mathbb O}({\bf v}_{\vertex}, 
\alpha \gamma)$ by 
\begin{eqnarray}
s_{\vertex} \left( [ r_1 e^{j \theta_1}, \hsppp 
r_2 e^{j \theta_2},  \cdots ,  r_{N_t} e^{j \theta_{N_t}} ]^{\sl T} 
\right) = 
\left[ \sqrt { 1 - \alpha^2(1 - r_1^2) } e^{j \theta_1}, \hsppp 
 \alpha \hsppp r_2 e^{j \theta_2},  \cdots , \alpha \hsppp r_{N_t} 
e^{j \theta_{N_t}} \right]^{\sl T} 
\end{eqnarray} 
where we have denoted the vector in the argument on the left side of the 
above equation in its polar form. The map $s_{\bef}(\cdot)$ defined as a 
composition $s_{\bef} = r_{\vertex}^{-1} \circ s_{\vertex} \circ r_{\vertex}$ 
results in 
\begin{eqnarray} 
s_{\bef}(\bfa_i) =  \bfa_1 \sqrt{1 - \alpha^2(1 - |\bfa_1^H \bfa_i|^2)} 
e^{j \angle{\bfa_1^H \bfa_i} } + \alpha \bfa_1^{\perp} 
\bfa_1^{\perp, \hsppp H} \bfa_i. 
\label{sbef}
\end{eqnarray} 
It can be checked that $s_{\bef}(\bfa_1) = \bfa_1$ on ${\cal G}(N_t,1)$. 
Furthermore, 
the inner product of the second term with $\bfa_1$ is zero. 
Hence, $d\left(s_{\bef}(\bfa_i), s_{\bef}(\bfa_1) \right) = 
d\left(s_{\bef}(\bfa_i), \bfa_1 \right) = \alpha d(\bfa_i, \bfa_1)$ for all $i$. 
\endproof 
\end{lem}

The rotation and scaling maps to be proposed now generalize the 
result of~\cite{samanta_heath} to the precoding scenario, $M > 1$. 
\begin{thm}
\label{rot_prp1}
Let ${\cal{R}} = \{ \bFa_i, i = 1, \cdots, N \}$ be a root codeset centered 
around $\bFa_1$ with a packing density $\gamma$. Let the $N_t \times M$ 
semiunitary matrix $\bFa_{\target}$ be the desired center of the rotated codeset. 
Then, the rotated codeset ${\cal{G}}$ is given by ${\cal{G}} = \{ \bG_i, i = 1, 
\cdots, N \}$ where $\bG_i = \bU_{\bFa_{\target}} \hspp \bU_{\bFa_1}^H \hspp \bFa_i$ 
with unitary matrices $\bU_{\bFa_1}$ and $\bU_{\bFa_{\target} }$ defined as 
$\bU_{\bFa_1} = \left[ \bFa_1 \hsp \bFa_1^{\nulla} \right]$ and 
$\bU_{\bFa_{\target} } = \left[ \bFa_{\target} \hsp \bFa_{\target}^{\nulla} 
\right]$. Here, $\bFa_1^{\nulla}$ and $\bFa_{\target}^{\nulla}$ are $N_t \times 
(N_t - M)$-dimensional representatives of the null-spaces of $\bFa_1$ and $\bFa_{\target}$, respectively. 
\end{thm}
\begin{proof}
See Appendix~\ref{app_rot}. 
\end{proof}

Note that there exists more than one basis for the null-space and therefore the 
usage of the term ``representative'' in the statement of the theorem. The lack of a 
unique representative for the null-space is responsible for the non-uniqueness of 
the rotation map that can effect a desired rotation. 

Before we get into the most general form of the scaling map, we illustrate 
a special case of it so as to provide insights into the construction. 
As before, let ${\cal{R}} = \{ \bFa_i, i = 1, \cdots, N \}$ be a root codeset 
centered around $\bFa_1$ with a packing density $\gamma$. Let ${\bf V}_1 = 
[{\bf v}_1  \cdots {\bf v}_M]$ where ${\bf v}_i$ is an $N_t \times 1$ vector 
and is the $i$-th column of ${\bf V}_1$. Define the map $s(\cdot)$ by 
\begin{eqnarray}
s({\bf V}_i) =  \left[ 
\begin{array}{ccccc}
{\bf v}_1 & {\bf v}_2 & \cdots & {\bf v}_{M-1 } & 
\beta {\bf v}_M + \delta {\bf v}_{M+1}
\end{array}
\right] 
\label{sdefn}
\end{eqnarray}
where $\beta = \sqrt{ 1 - \alpha^2 
\left(1 - \lambda_{\min} \left( {\bf V}_1^H {\bf V}_i {\bf V}_i^H {\bf V}_1 \right) 
\right) }$, $\delta = \alpha \sqrt{1 - \lambda_{\min} \left( 
{\bf V}_1^H {\bf V}_i {\bf V}_i^H {\bf V}_1 \right) }$, and ${\bf v}_{M+1}$ is 
orthogonal to ${\bf V}_1$ (that is, ${\bf v}_{M+1}^H {\bf V}_1 = {\bf 0}_{1 \times M}$). 
We illustrate three properties satisfied by $s(\cdot)$ which ensures that it can 
scale submanifolds. Noting that ${\bf v}_i, i = 1, \cdots , M+1$ are orthonormal 
vectors in ${\mathbb C}^{N_t}$ and that $\beta^2 + \delta^2 = 1$, it is straightforward 
to check that $s({\bf V}_i)^H s({\bf V}_i) = \bI_M$. For $s({\bf V}_1)$, note that 
$\beta = 1$ and $\delta = 0$ which results in $s({\bf V}_1) = {\bf V}_1$. 

\begin{prop}
\label{lem_splcase}
We also have $d(s({\bf V}_1) , s({\bf V}_i)  ) = \alpha d({\bf V}_1, {\bf V}_i)$ 
for any $i \neq 1$. Thus, $s(\cdot)$ induces the scaling of ${\cal R}$ by $\alpha$. 
\end{prop}
\noindent 
\begin{eqnarray}
{\it Proof:} \hspp {\rm Note} \hspp {\rm that} \hspp 
d(s({\bf V}_1), s({\bf V}_i) ) & \stackrel{(a)}{=} & d({\bf V}_1 , s({\bf V}_i)) = 
\lambda_{\max}( {\bf V}_1 {\bf V}_1^H - s({\bf V}_i) s({\bf V}_i)^H  ) 
\nonumber \\ 
& \stackrel{(b)}{=} &
\lambda_{\max}(  {\bf v}_M {\bf v}_M^H - 
( \beta {\bf v}_M + \delta {\bf v}_{M+1} )  
( \beta {\bf v}_M + \delta {\bf v}_{M+1} )^H ) 
\end{eqnarray}
where in (a) we have used $s({\bf V}_1) = {\bf V}_1$ and (b) follows from~(\ref{sdefn}). 
Using the trick of Lemma~\ref{lem_dist}, observe that the square of $\lambda_{\max}$ 
in the above equation satisfies 
$\lambda_{\max}^2 = 1 - | {\bf v}_M^H (\beta {\bf v}_M + \delta {\bf v}_{M+1})  |^2 
= 1 - \beta^2 = \alpha^2 (1 - \lambda_{\min}( {\bf V}_1^H {\bf V}_i {\bf V}_i^H {\bf V}_1 ))$. 
The proof is complete by noting the value of $d({\bf V}_1, {\bf V}_i)$ from 
Prop.~\ref{prop_000}. 
\endproof 

The choice of ${\bf v}_{M+1}$ is not unique 
and it is not clear whether the map in~(\ref{sdefn}) is unique modulo 
the choice of ${\bf v}_{M+1}$. Furthermore, note that when $(N_t - M) \geq M$, 
$s({\bf V}_i)$ can be written as 
\begin{eqnarray}
s({\bf V}_i) = {\bf V}_1 {\bf A}_i + {\bf V}_1^{\nulla} {\bf B}_i 
\label{idea}
\end{eqnarray}
where ${\bf A}_i = \diag( \left[1 , \cdots , 1, \beta  \right] )$ and 
${\bf B}_i$ has only one non-zero entry which is at the $(M,M)$-th location and 
its value 
is $\delta$. In Appendix~\ref{app_scale}, we resolve 
the uniqueness issue and construct the most general form of $s(\cdot)$. We also show 
that the most general form of $s({\bf V}_i)$ is of the form in~(\ref{idea}) for a 
suitable choice of ${\bf A}_i$ and ${\bf B}_i$. 

\subsection{Reduction to the Beamforming Construction of Lemma~\ref{lem_scale}} 
\begin{cor} 
In the special case of $M = 1$, the scaling map proposed in~(\ref{sdefn}) 
(and extended in Theorem~\ref{scale_prp2} of Appendix~\ref{app_scale}) 
is a generalization of the map proposed in Lemma~\ref{lem_scale} (see~(\ref{sbef})). 
\label{cor_gen_sbef}
\end{cor} 
\begin{proof} 
For the sake of simplicity, we denote the map constructed in~(\ref{sdefn}) 
as $s_{\gen}(\cdot)$. We write $s_{\gen}(\cdot)$ as $s_{\gen}({\bf v}_i) = 
\sqrt{\lambda_i} {\bf v}_1 + \sqrt{1 - \lambda_i} {\bf v}_1^{\nulla}$ where 
$\lambda_i = 1 - \alpha^2 (1 - |{\bf v}_1^H {\bf v}_i|^2)$ and ${\bf v}_1^{\nulla}$ 
is an $N_t \times 1$ unit norm vector orthogonal to ${\bf v}_1$. We now draw a 
correspondence between $s_{\bef}(\cdot)$ and $s_{\gen}(\cdot)$. 

In Lemma~\ref{lem_scale}, note that ${\bf U}_{\vertex} \bU_{\vertex}^H = \bI_{N_t}$ 
which implies that $\bfa_1^{\perp} \bfa_1^{\perp,\hsppp H} = \bI_{N_t - 1}$. Using the 
fact that ${\bf U}_{\vertex}^H \bU_{\vertex} = \bI_{N_t}$, similarly we obtain 
$\bfa_1^{\perp,\hsppp H} \bfa_1^{\perp} = \bI_{N_t } - {\bf v}_1 {\bf v}_1^H$. 
Using this in~(\ref{sbef}), we have 
\begin{eqnarray}
s_{\bef}(\bfa_i) & = & \sqrt{1 - \alpha^2(1 - |\bfa_1^H \bfa_i|^2)} 
e^{j \angle{\bfa_1^H \bfa_i} } \bfa_1
+ \alpha (\bfa_i - \bfa_1( \bfa_1^H \bfa_i  )) 
\\ & = & \sqrt{\lambda_i} e^{j \angle{\bfa_1^H \bfa_i} } \bfa_1
+ \alpha (\bfa_i - \bfa_1( \bfa_1^H \bfa_i  )). 
\end{eqnarray}
It is straightforward but surprising to note that 
$\frac{\bfa_i - \bfa_1( \bfa_1^H \bfa_i  )} {\sqrt{1 - | \bfa_i^H \bfa_1|^2 } }$ 
is both unit norm and orthogonal to $\bfa_1$. Further, note that 
$\sqrt{1 - \lambda_i} = \alpha \sqrt{1 - | \bfa_i^H \bfa_1|^2 }$. By setting 
$\frac{\bfa_i - \bfa_1( \bfa_1^H \bfa_i  )} {\sqrt{1 - | \bfa_i^H \bfa_1|^2 } }$ as 
the representative of ${\bf v}_1^{\nulla}$ in the general framework, we 
see that $s_{\bef}(\cdot)$ can be obtained up to a phase term. And since we 
operate on the Grassmann manifold which is impervious to right multiplication by 
terms of the form $e^{j \theta}$, we have proved the corollary. 
\end{proof}

\subsection{Low-Complexity Generation of Local Components} 
We now illustrate how the theory of rotation and scaling maps can be used to 
construct precoding codebooks with low-complexity. 

\noindent{\bf{\emph{Root Codeset Generation:}}} 
A root codeset that satisfies the localization and well-packing conditions as 
described above is constructed via VQ. The number of codewords in the root codeset 
is larger than $N_{\loc}^1$ so as to ensure that any local component has a 
cardinality smaller than that of the root codeset. Furthermore, since the scaling 
map can only ensure that the output packing is more localized than the input packing, 
we need to pick $\theta$ sufficiently large, but smaller than $1$. The quantity 
$\gamma_{\max}(N_t, M, N, \theta)$ corresponding to the choices of $N_t, M, N$ and 
$\theta$ is determined via Monte Carlo techniques and some $\gamma$ is chosen in the 
interval $(0, \gamma_{\max}(N_t, M, N, \theta) )$. 

\noindent{\bf{\emph{Local Components:}}} 
For each member of the statistical component, we rotate the root codeset (via the 
rotation map of Theorem~\ref{rot_prp1}) to the $N_t \times M$ matrix corresponding 
to the subspace of ${\bf \Sigma}_t$ in the statistical component. Then, each rotated 
codeset is scaled by a shrinking factor $\alpha_i \triangleq \frac{\mu_i}{\mu_1}$. 
That is, we scale each rotated codeset in proportion to the generalized eigenvalue 
of that subspace. From each rotated codeset of $N$ codewords, we retain 
$N_{\loc}^i, i = 1, \cdots, N_{\stat}$ codewords. The heuristic behind the choice of 
$N_{\loc}^i$ has been explained in the previous section. The same heuristic can be 
used to justify the choice of $\alpha_i$ as well.

\subsection{Exploiting the General Structure of the Scaling and Rotation Maps} 
We now delve into why a general form of the maps in Appendix~\ref{app_scale} is 
useful. In many practical systems, it is desired that the precoder codebook has 
more structure so as to ensure implementation ease. For example, two commonly 
desired properties are: 
\newline \noindent 1) {\em Bounded Gain Power Amplifier Architecture} where we 
require 
\begin{eqnarray}
\max_{ {\bf V}_i \hsppp \in \hsppp {\cal C} } 
\frac{  \max_{mn} | {\bf V}_i(m,n) | } {\min_{mn} |{\bf V}_i(m,n)| } 
\leq \eta. 
\end{eqnarray} 
The above condition is useful in ensuring that the power amplifiers used in the radio 
link chain are not driven to their operational limits. The most general form of the 
rotation and scaling maps allows one to search for a codebook that satisfies 
the above property in addition to the localization and well-packing properties, 
and 
\newline \noindent 2) {\em Recursive Codebook Structure} where a codebook of 
${\sf rank}$-$N_{\sf small}$ can be generated from a codebook of 
${\sf rank}$-$N_{\sf large}$ (with $N_{\sf large} > N_{\sf small}$) by retaining 
only a subset of $N_{\sf small}$ columns from every precoder in the 
${\sf rank}$-$N_{\sf large}$ codebook. This property is desired so as to 
minimize the algorithmic complexity of generating a family of codebooks of different 
ranks {\em on the fly}. The low-complexity property of the proposed maps and the 
{\em offline} generation of the root codesets of different ranks ensure that this issue 
is redundant with our codebook design. 

Thus, we strongly generalize the maps of~\cite{samanta_heath} and as a by-product 
observe that even in the $M = 1$ case, a rich family of maps can effect the 
scaling operation other than~(\ref{sbef}). Additional structure in the codebook 
can also be accommodated to ease implementation complexity.

\begin{figure}[htb!]
\begin{center}
\begin{tabular}{cc}
\includegraphics[width=3.2in]{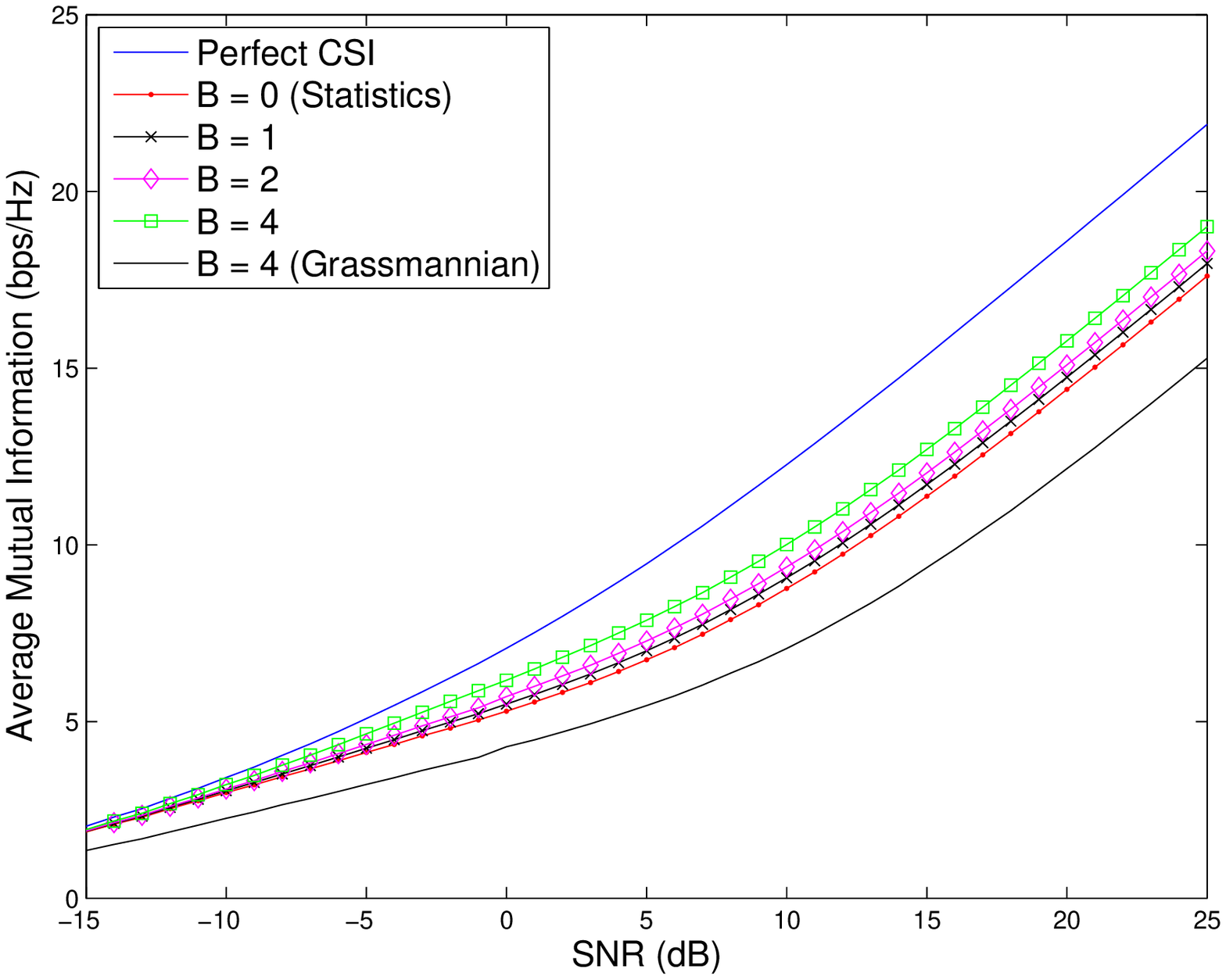} & 
\includegraphics[width=3.2in]{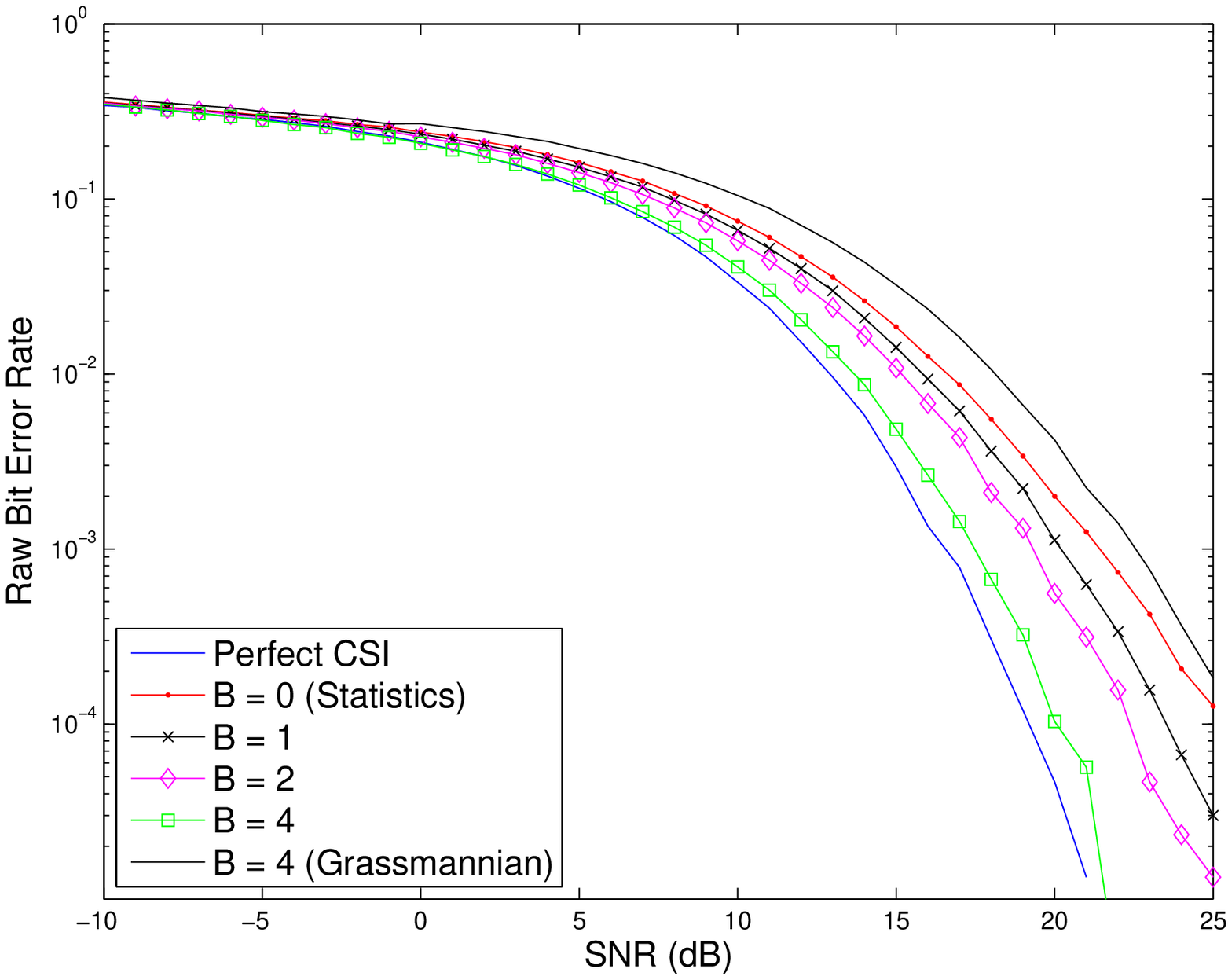} 
\\ (a) & (b) 
\end{tabular} 
\caption{\label{fig_442kron}(a) Average mutual information with Gaussian 
inputs in a $4 \times 4$ {\em mismatched} channel following a separable model. 
Two data-streams are used in signaling and a limited feedback codebook designed 
along the principle elucidated in Sec.~\ref{sec4} is used. (b) Error probability 
performance with the same codebook under QPSK inputs.} 
\end{center}
\end{figure}



\section{Numerical Results}
\label{sec5} 
We now illustrate via numerical studies the performance gains possible with our 
codebook construction and the consequent bridging of the gap between statistical 
and optimal precoding. In the first study, we consider a $4 \times 4$ channel 
under the separable model with ${\bf \Lambda}_t = 
\diag \left([14.98 \hsp 0.50 \hsp 0.26 \hsp 0.26] \right)$ and 
${\bf \Lambda}_r = \diag \left( [ 15.5 \hsp 0.25 \hsp 0.15 \hsp 0.10 ] \right)$. 
This choice ensures that the transmit$/$receive covariance matrices are both 
ill-conditioned and with $M = 2$, note that the channel is {\em not} matched to the 
precoder. We 
first generate a root codeset of $N = 4$ codewords with $\theta \approx 0.76$ and 
$\gamma \approx 0.75$ via VQ. Let $\{\bu_i \}$ be the column vectors of $\bU_t$. 
The codebook used for $B = 1$ satisfies $N_{\stat} = 1$ with the codeword corresponding 
to $[\bu_1 \hspp \bu_2]$ and $N_{\glo} = 1$ while with $B = 2$, the codebook has an 
additional RVQ codeword and a local codeword around $[\bu_1 \hspp \bu_2]$. Similarly, 
with $B = 4$, $N_{\stat} = 3, N_{\loc}^1 = N_{\loc}^2 = 3, N_{\loc}^3 = 2$ and 
$N_{\glo} = 5$. The statistical codewords correspond to $[\bu_1 \hspp \bu_i], \hspp 
i = 2, \cdots , 4$. Since we are mainly interested in illustrating the performance 
gains in the high-$\snr$ regime, uniform power allocation is used for 
${\bf \Lambda}_{\stat}$.

Fig.~\ref{fig_442kron}(a) shows the average mutual information 
with a Gaussian input for statistical and limited feedback precoding. In addition to 
the mutual information, raw bit error rate (BER) is useful as well. 
Fig~\ref{fig_442kron}(b) shows the improvement in error probability in the same 
channel with QPSK inputs. In the error probability case, the index of the codeword 
that minimizes the distance to the instantaneous $\widetilde{\bf V}_{\bf H}$ is fed 
back. Note that while the performance gap between the optimal 
and the statistical schemes is significantly bridged in the error probability case, 
further improvement in mutual information is possible. Nevertheless, both the 
figures show that substantial gains are possible with a few bits of feedback. For 
example, with $B = 4$ bits of feedback, a $3$ dB gain is possible at a rate of 
$10$ bps/Hz while a $6$ dB gain is possible at a BER of $10^{-3}$. Also, note that an 
i.i.d.\ codebook design incurs a dramatic loss in performance in correlated channels.

\begin{figure}[htb!]
\centering
\begin{tabular}{c}
\includegraphics[height=3.2in,width=4in]{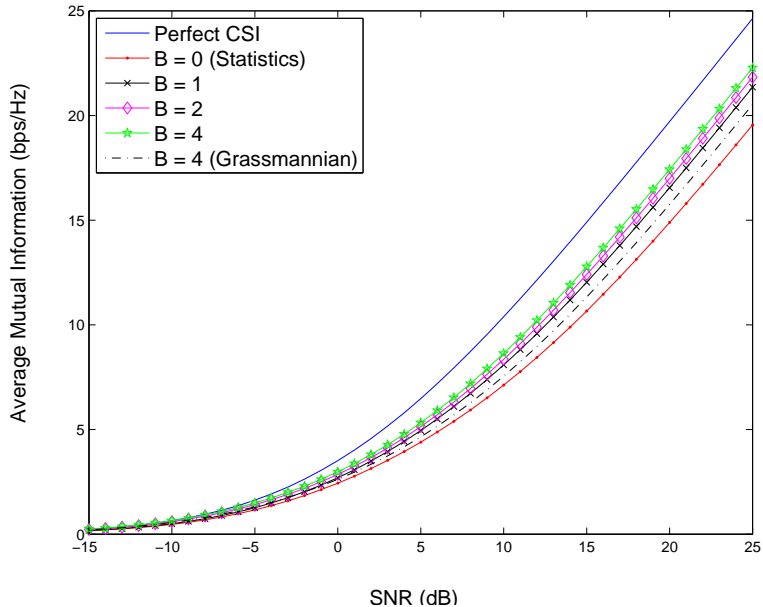} 
\end{tabular}
\caption{Average mutual information with Gaussian inputs in a $4 \times 4$ 
{\em mismatched} channel with non-separable correlation and $M = 3$.} 
\label{fig_443can} 
\end{figure}

In the second study, we consider a $4 \times 4$ channel with non-separable 
correlation following the virtual representation framework. The variance matrix 
${\bf \sigma}(i,j) \triangleq \bEe \left[ |\bH_{\ind}(i,j)|^2 \right]$ 
used in the study is 
\begin{eqnarray}
{\bf \sigma} = \left [ \begin{array}{cccc} 
1.24 & 1.42 & 7.49 & 0.23 \\ 
0.41 & 0.14 & 0.42 & 0.03 \\ 
0.72 & 1.39 & 0.07 & 0.02 \\ 
0.28 & 0.13 & 0.50 & 1.51 \end{array}
\right]. 
\end{eqnarray} 
\newline 
Note that the channel has a single dominant transmit (as well as receive) 
eigen-mode and is hence mismatched when $M = 3$ data-streams are used in signaling. 
The parameters of the root codeset are $N = 4, \hspp \theta \approx 0.87$ and 
$\gamma \approx 0.84$. As before, let $\{\bu_i \}$ be the column vectors of the DFT 
matrix $\bU_t$. The codebook 
for $B = 1$ has the two statistical codewords $[ \bu_3 \hspp \bu_2 \hspp \bu_1 ]$ 
and $[ \bu_3 \hspp \bu_2 \hspp \bu_4 ]$. For $B = 2$, we use two additional RVQ 
codewords and for $B = 4$, we use $N_{\stat} = 3 , N_{\loc}^1 = N_{\loc}^2 = 3, 
N_{\loc}^3 = 2$ and $N_{\glo} = 5$. The third statistical codeword when $B = 4$ 
is $[\bu_3 \hspp \bu_1 \hspp \bu_4]$. Fig.~\ref{fig_443can} illustrates the 
bridging of the gap in mutual information between the optimal and the statistical 
schemes. It is important to note that both the channels studied here are so constructed 
to result in a substantial performance gap between perfect CSI and statistical signaling. 
In more realistic channels that are not so poorly matched, we expect an even 
better performance with our scheme. Thus our studies illustrate that substantial 
gains can be achieved even with few bits of feedback.

\section{Concluding Remarks} 
\label{sec6} 
In this work, we have studied linear precoding under a realistic system model. 
In particular, the focus is on the impact of spatial correlation when perfect CSI 
is available at the receiver, statistical information is available at both the ends, 
and quantized channel information is fed back from the receiver to the transmitter. 
While initial works on precoding assume perfect CSI at both the ends and hence do 
not impose any particular structure on the precoder matrices, under the model 
studied here, we see that structure can help in minimizing the reverse-link feedback 
as well as ease the implementation complexity. 

We introduced the notion of matched and mismatched channels and illustrated that 
limited feedback precoding is useful only in the case of mismatched channels. The 
study of statistical precoding motivates the 
proposed limited feedback design where we quantize the space of semiunitary matrices 
with a non-uniform bias towards the statistically dominant eigen-modes. The design as 
well as its adaptability are rendered practical by the construction of mathematical 
maps (operations) that can rotate and scale submanifolds on the Grassmann manifold. 
More importantly, numerical studies show that the proposed designs 
yield significant improvement in performance when the channel is mismatched to the 
communication scheme. 

This work is a first attempt at systematic precoder codebook design in 
single-user multi-antenna channels that exploits spatial correlation. Possible 
extensions are the study of more complex receiver architectures and performance 
analysis in the finite antenna, arbitrary $\snr$ setting, along the lines 
of~\cite{stat_semiunitary}. 
More work also needs to be 
done to understand the impact of spatial correlation on the performance of the 
proposed limited feedback scheme which could in turn drive the development of 
more efficient codebook constructions. Open issues that need further study 
include practical aspects like codebook designs for wideband channels, 
codebook designs based on Fourier$/$Hadamard matrices that are useful in achieving 
the bounded gain power amplifier architecture and hence, have found much interest 
in the standardization community, incorporating the cost of statistics 
acquisition in performance analysis~\cite{vasanth_icassp08}, 
and more general scattering environment-independent channel 
decompositions~\cite{vasanth_asilomar} 
that mimic the physical model closely. The case of multi-user systems with 
feedback, which has attracted significant recent interest, is another area for study. 

We close the paper by drawing attention to the philosophy that has guided this work. 
While deducing the structure of the optimal signaling scheme under general assumptions 
on spatial correlation and channel information seems extremely difficult, an 
alternative approach that partitions this problem into smaller sub-problems could 
be quite fruitful. The general idea of matching the rank of the precoding scheme 
to the number of dominant transmit eigenvalues with the resolution necessary to 
decide whether an eigenvalue is ``dominant'' or not being a function of the $\snr$ 
reminds one of the classical source-channel matching paradigm~\cite{cover_thomas}. 
Initial evidence seen in this paper also 
suggests that this partitioning provides a natural framework to understand the 
performance of limited feedback schemes.

\appendix 

\ignore{ 
\subsection{Proof of Prop.~\ref{prop_000}} 
\label{app_000}
The facts that $\bFa_1 \bFa_1^H - \bFa_2 \bFa_2^H$ is Hermitian and its trace 
equals zero implies that $\lambda_{\max}( \bFa_1 \bFa_1^H - \bFa_2 \bFa_2^H ) < 0$ 
is impossible. 
If $d_{\proj, \hsppp 2}(\bFa_1, 
\bFa_2) = 0$, then it is easy to see that $\bFa_1 \bFa_1^H = \bFa_2 \bFa_2^H$ 
from which we note that $\bFa_1^H \bFa_2 = {\bf U}$ for some $M \times M$ unitary 
${\bf U}$ (which is equivalent to the fact ${\bf V}_1 = {\bf V}_2$ on 
${\cal G}(N_t, M)$). The other direction follows trivially. 
For the upper bound, we the need the following result~\cite{silvester} that 
helps in computing the determinant of partitioned matrices. 
\begin{lem} 
\label{lem_partition}
If ${\bf X}, {\bf Y}, {\bf Z}$ and ${\bf W}$ are $n \times n$ matrices and ${\bf W}$ 
is invertible, we have 
\begin{eqnarray} 
\det \left[ \begin{array}{cc}
{\bf X} & {\bf Y} \\ 
{\bf Z} & {\bf W} 
\end{array}  \right] = \det({\bf X} - {\bf Y} {\bf W}^{-1} {\bf Z}) 
\cdot \det({\bf W}). 
\end{eqnarray}
\endproof 
\end{lem} 
Using the above fact, it is straightforward to check that $d_{\proj, \hsppp 2} 
(\bFa_1, \bFa_2) = \sqrt{1 - \lambda_{\min}( \bFa_1^H \bFa_2 \bFa_2^H \bFa_1  )}$ 
which results in the proposition. 
\endproof 
}

\subsection{Proof of Theorem~\ref{rot_prp1}}
\label{app_rot}
\ignore{ 
\noindent {\bf{\emph{Preliminaries:}}} 
The following lemma~\cite{silvester} helps in computing the determinant of 
partitioned matrices. 
\begin{lem} 
\label{lem_partition}
If ${\bf X}, {\bf Y}, {\bf Z}$ and ${\bf W}$ are $n \times n$ matrices and ${\bf W}$ 
is invertible, we have 
\begin{eqnarray} 
\det \left[ \begin{array}{cc}
{\bf X} & {\bf Y} \\ 
{\bf Z} & {\bf W} 
\end{array}  \right] = \det({\bf X} - {\bf Y} {\bf W}^{-1} {\bf Z}) 
\cdot \det({\bf W}). 
\end{eqnarray}
\endproof 
\end{lem} 
}
The efficacy of the rotation map is established if we can show the following: 
\begin{enumerate}
\item 
$\bG_i^H\bG_i = \bI_M$ for all $i$, 
\item 
$\bG_1 = \bFa_{\target}$, and 
\item 
$d_{\proj,\hsppp 2}(\bG_1 , \bG_i) = d_{\proj,\hsppp 2}(\bFa_1 ,\bFa_i)$ 
for all $i$. 
\end{enumerate} 

To prove 1), first note that $\bU_{\bFa_1}$ and $\bU_{\bFa_{\target}}$ are 
$N_t \times N_t$ unitary matrices. From the semiunitarity property of 
$\bFa_i$, $\bG_i^H \bG_i = \bI_M$ follows trivially. Using the unitary property 
of $\bU_{\bFa_1}$ and the decomposition in the statement of the theorem, 
2) also follows trivially. For 3), note that 
\begin{eqnarray}
d_{\proj,\hsppp 2}(\bG_1,\bG_i)  
& = & \lambda_{\max}\left( \bG_1 \bG_1^H - \bG_i \bG_i^H \right) \nonumber \\ 
& = & \lambda_{\max} \left( \bU_{\bFa_{\target}} \bU_{\bFa_1}^H 
\left(\bFa_1\bFa_1^H - \bFa_i \bFa_i^H\right) 
\bU_{\bFa_1} \bU_{\bFa_{\target}}^H \right) 
= d_{\proj,\hsppp 2}(\bFa_1 , \bFa_i). 
\end{eqnarray}
In the above chain of equalities, we have used the fact that $\lambda({\bf AB}) = 
\lambda({\bf BA})$ and the unitary property of $\bU_{\bFa_1}$ and 
$\bU_{\bFa_{\target}}$. Thus the proof is complete. 
\endproof


\subsection{Generalized Scaling Map}
\label{app_scale}
\begin{thm} 
\label{scale_prp2}
Let ${\cal{R}}$ be a root codeset with packing density $\gamma$ and center $\bFa_1$. 
Let $\bU_{\bA}$ and ${\bf W}$ be arbitrary $M \times M$ unitary matrices and let 
$\bU_{\bB}$ be an arbitrary $(N_t - M) \times (N_t - M)$ unitary matrix. 
Given $\alpha \in (0,1)$ and $M \leq (N_t - M)$, for any $\bFa_i \in {\cal{R}}$, 
generate an $M \times M$ diagonal, positive-definite matrix $\bfLambda_{i}$ 
with: $\bfLambda_{\min} \triangleq 
\min_j \bfLambda_i(j) = 1 - \alpha^2 \left( 1 - \lambda_{\min}(\bFa_1^H 
\bFa_i \bFa_i^H \bFa_1) \right)$ and $\bfLambda_{\max} \triangleq 
\max_j \bfLambda_i(j) \leq 1$. Then, define $\bA_i$ as 
$\bA_i = \bU_{\bA} \hsppp \bfLambda_i^{1/2} \hsppp {\bf W}^H$. 
Define the $M \times M$ principal component of the $(N_t - M) \times M$ diagonal 
matrix ${\bfLambda}_{\bB}$ as $\left( \bI_M - \bfLambda_i \right)^{1/2}$ 
and $\bB_i$ as 
$\bB_i = \bU_{\bB} \hsppp \bfLambda_{\bB}^{1/2} \hsppp {\bf W}^H$. 

If $M > (N_t - M)$, for any $\bFa_i \in {\cal{R}}$, generate an $(N_t - M) 
\times (N_t - M)$ diagonal, positive-semidefinite matrix ${\bf \Gamma}_{i}$ 
with: ${\bf \Gamma}_{\max} \triangleq 
\max_j {\bf \Gamma}_i(j) = \alpha^2 \left( 1 - \lambda_{\min}(\bFa_1^H 
\bFa_i \bFa_i^H \bFa_1) \right)$ and ${\bf \Gamma}_{\min} \triangleq 
\min_j {\bf \Gamma}_i(j) \geq 0$. Then, define $\bB_i$ as $\bU_{\bB} \hsppp 
\bfLambda_{\bB}^{1/2} \hsppp {\bf W}^H$ 
with the principal $(N_t - M) \times (N_t - M)$ 
component of $\bfLambda_{\bB}$ being ${\bf \Gamma}_i$. Define ${\bf A}_i$ as 
$\bA_i = \bU_{\bA} \hsppp \bfLambda_{\bA} \hsppp {\bf W}^H$ with the principal 
$(N_t -  M) \times (N_t - M)$ component of $\bfLambda_{\bA}$ being 
$\bI_{N_t - M} -{\bf \Gamma} _i $ and the principal 
southeast component being $\bI_{2M - N_t}$. 

Then, the scaling map $s(\cdot)$ that leads to a packing density of $\gamma \alpha$ 
is given by 
\begin{eqnarray}
\label{eqna}
s(\bFa_i) = \bFa_1 \hsppp \bA_i + \bFa_1^{\nulla} \hsppp \bB_i 
\end{eqnarray}
where $\bFa_1^{\nulla}$ is a representative of the null-space corresponding to $\bFa_1$.
\end{thm}
\ignore{ 
where $\bfLambda_i$ is an $M \times M$ positive-semidefinite diagonal 
matrix. Also, define the $(N_t - M) \times M$ matrix $\bB_i$ as 
\begin{eqnarray}
\bB_i = \left\{ 
\begin{array}{l} 
\bU_{\bB} \left[ 
\begin{array}{cc}  
\left( \bI_{N_{\min}} -  \bfLambda_{i, \hsppp 0} \right)^{1/2}  & 
\bO_{N_{\min} \times (N_{\max} - N_{\min}) } 
\end{array} \right] \bV^H 
{\hspace{0.2in}} 
{\mathrm{if}} \hspp N_t - M < M \\ \\ 
\bU_{\bB} \left[ 
\begin{array}{c}
\left( \bI_{N_{\min}} - \bfLambda_{i, \hsppp 0} \right)^{1/2} \\ 
\bO_{N_{\max} - N_{\min} \times N_{\min}}
\end{array}
\right] \bV^H 
{\hspace{0.2in}} {\mathrm{if}} \hspp N_t - M \geq M 
\end{array} 
\right. 
\nonumber 
\end{eqnarray}
where $\bfLambda_{i, \hsppp 0}$ is the $N_{\min} \times N_{\min}$ principal 
submatrix of $\bfLambda_i$. 

The constraints on $\bfLambda_i$ are: $\bfLambda_{\min} \triangleq 
\min_j( \bfLambda_i(j)) = 1 - \alpha^2 \left( 1 - \lambda_{\min}(\bFa_1^H 
\bFa_i \bFa_i^H \bFa_1) \right)$ and $\bfLambda_{\max} \triangleq \max_j (\bfLambda_i(j)) 
\leq 1$ with $\bfLambda_i(j) = 1$ for $j > (N_t - M)$ to $M$ if $(N_t - M) < M$. 

Then, the scaling map $s(\cdot)$ that leads to a packing density of $\gamma \alpha$ 
is defined as 
\begin{eqnarray}
\label{eqna}
s(\bFa_i) = \bFa_1 \hsppp \bA_i + \bFa_1^{\nulla} \hsppp \bB_i 
\end{eqnarray}
where $\bFa_1^{\nulla}$ is a representative of the null-space corresponding to $\bFa_1$.
\end{thm}
}
\begin{proof}
Let $r_{\bU}$ denote the rotation effected by a unitary matrix $\bU$. Since the 
scaling operation has to keep the center of a root codeset fixed, in the sequel, 
we use a fixed $N_t \times M$ matrix as the center instead of $\bFa_1$ which is 
dependent on the choice of ${\cal{R}}$. This is achieved by a composition of three 
operations: 
\begin{eqnarray} 
\label{map} 
s(\cdot) = r_{\bU_{\bFa_1}} \odot s_{\vertex} \odot r_{\bU_{\bFa_1}^H} (\cdot). 
\end{eqnarray}
Here, $r_{\bU_{\bFa_1}^H}(\cdot)$ rotates the root codeset to the canonical precoder 
$[\bI_M \hspp \bO_{M \times (N_t - M)}]^T$ while $s_{\vertex}(\cdot)$ scales 
(shrinks) the canonical codeset by a factor $\alpha$ and $r_{\bU_{\bFa_1}}$ rotates 
it back to the direction corresponding to $\bFa_1$. From the above definition of 
$s(\cdot)$, we have 
\begin{eqnarray}
s(\bFa_i) = \left[ 
\begin{array}{cc} 
\bFa_1 & \bFa_1^{\nulla} 
\end{array} \right] \hspp 
s_{\vertex} \left( \left[ 
\begin{array}{c} 
\bFa_1^H \hsppp \bFa_i \\ 
\bFa_1^{\nulla,H} \hsppp \bFa_i 
\end{array} \right] \right) 
= \left[ \begin{array}{cc} 
\bFa_1 & \bFa_1^{\nulla} 
\end{array} \right] 
\hspp \left[ 
\begin{array}{c} 
\bA_i \\ \bB_i \end{array} \right] 
= {\bf V}_1 {\bf A}_i + {\bf V}_1^{\nulla} {\bf B}_i 
\nonumber 
\end{eqnarray} 
where we have used a partitioning $[\bA_i^T  {\hspace{0.05in}} \bB_i^T]^T$ for the 
$N_t \times M$ matrix 
$s_{\vertex} \left( \left[ \begin{array}{c} \bFa_1^H \hsppp \bFa_i \\ 
\bFa_1^{\nulla,H} \hsppp \bFa_i \end{array} \right] \right)$. 
In this partitioning, $\bA_i$ is $M \times M$ and is of full rank while 
$\bB_i$ is an $(N_t - M) \times M$ matrix. 

Given that $\bFa_1^H\bFa_1 = \bI_M$, $\bFa_1^H \bFa_1^{\nulla} = 
\bO_{M \times (N_t - M)}$ and $\bFa_1^{\nulla,H} \bFa_1^{\nulla} = \bI_{N_t - M}$, 
the relationship $\bA_i^H \bA_i + \bB_i^H \bB_i = \bI_M$ ensures that $s({\bf V}_i)$ 
is semiunitary. We show that $\bA_i$ and $\bB_i$ have to be as in the statement of 
the theorem so that the following properties of $s(\cdot)$ are met: 
\begin{enumerate}
\item 
$d_{\proj, \hsppp 2}( s({\bf V}_1), s({\bf V}_i)  ) = \alpha d_{\proj, \hsppp 2}
({\bf V}_1, {\bf V}_i)$ for all $i$, and 
\item 
$s({\bf V}_1) = {\bf V}_1$. 
\end{enumerate}

First, let us consider the distance scaling property. 
Assuming 2) (which we check subsequently) and following Prop.~\ref{prop_000}, we need 
\begin{eqnarray} 
\lambda_{\max} (\bC) = 
\lambda_{\max} \left( \bFa_1 \bFa_1^H - s(\bFa_i) s(\bFa_i)^H  \right) 
= \alpha \sqrt{1 - \lambda_{\min} (\bFa_1^H \bFa_i \bFa_i^H \bFa_1 ) }  
 \label{eqnb}
\end{eqnarray}
where $\bC \triangleq 
\bFa_1 \bFa_1^H - \bFa_1 \bA_i \bA_i^H \bFa_1^H - \bFa_1^{\nulla} 
\bB_i \bA_i^H \bFa_1^H - \bFa_1 \bA_i \bB_i^H \bFa_1^{\nulla,H} - 
\bFa_1^{\nulla} \bB_i \bB_i^H \bFa_1^{\nulla,H}$. In the expansion for $\bC$, we have 
used the relationship in~(\ref{eqna}). We can decompose $\bC$ as $\bC_2 \bC_1^T$ 
where 
\begin{eqnarray}
\begin{array}{cc} 
{\bC_1^T  = 
\left[ \begin{array}{c} 
\bFa_1^H \\ 
\bA_i^H \bFa_1^H \\ 
\bB_i^H \bFa_1^{\nulla,H} \\
\bB_i^H \bFa_1^{\nulla,H} 
\end{array} \right], } & 
{\bC_2 =  \left[ \begin{array}{cccc}
\bFa_1(\bI_M - \bA_i \bA_i^H) & - \bFa_{1}^{\nulla} \bB_i & - \bFa_1 \bA_i & 
-\bFa_1^{\nulla} \bB_i \end{array} \right]}. 
\end{array}
\end{eqnarray}
Note that the non-trivial eigenvalues of ${\bf A}{\bf B}$ are the same as 
those of ${\bf B}{\bf A}$. Hence, 
$\lambda_{\max}(\bC) = \lambda_{\max}(\bC_1^T \bC_2)$. Using the facts 
$\bFa_1^H\bFa_1 = \bI_M$, $\bFa_1^H \bFa_1^{\nulla} = \bO_{M \times (N_t - M)}$ 
and $\bFa_1^{\nulla,H} \bFa_1^{\nulla} = \bI_{N_t - M}$, observe that the 
$4M \times 4M$ matrix $\bC_1^T \bC_2$ is given by 
\begin{eqnarray}
\bC_1^T \bC_2 = \left[ \begin{array}{cccc}
\bI_M - \bA_i \bA_i^H & \bO_M & -\bA_i & \bO_M \\ 
\bA_i^H(\bI_M - \bA_i \bA_i^H ) & \bO_M & -\bA_i^H \bA_i & \bO_M \\ 
\bO_M & -\bB_i^H \bB_i & \bO_M & -\bB_i^H\bB_i \\ 
\bO_M & -\bB_i^H \bB_i & \bO_M & -\bB_i^H\bB_i \\ 
\end{array} \right]. 
\end{eqnarray}

We will now show that the largest eigenvalue of $\bC_1^T \bC_2$ can be computed 
in closed-form. For this, we need to solve for $\lambda$ by setting 
$\det(\bC_1^T \bC_2 - \lambda\bI_{4M}) = 0$. Towards this computation, we need 
to use Lemma~\ref{lem_partition} following which, we have 
\begin{eqnarray} 
\frac{ \det ( \bC_1^T \bC_2 - \lambda \bI_{4M}) }
{ \det(-\bB_i^H \bB_i - \lambda \bI_M) } = 
\det \left( \begin{array}{ccc} 
\bI_M - \bA_i \bA_i^H - \lambda \bI_M & \bO_M & -\bA_i \\ 
\bA_i^H(\bI_M - \bA_i \bA_i^H ) & - \lambda\bI_M & -\bA_i^H \bA_i \\ 
\bO_M & -\lambda \bB_i^H \bB_i (\bB_i^H \bB_i + \lambda \bI_M)^{-1} 
& -\lambda\bI_M  \\ 
\end{array} \right). 
\end{eqnarray}
With $\kappa = \det(-\bB_i^H \bB_i - \lambda \bI_M) \det(-\lambda \bI_M)$, 
upon another application of Lemma~\ref{lem_partition} we have 
\begin{eqnarray} 
\frac{ \det ( \bC_1^T \bC_2 - \lambda \bI_{4M}) }
{ \kappa } = \det \left( \begin{array}{cc} 
\bI_M - \bA_i \bA_i^H - \lambda \bI_M & \bA_i \bB_i^H \bB_i (\bB_i^H \bB_i 
+ \lambda \bI_M)^{-1} \\ 
\bA_i^H (\bI - \bA_i \bA_i^H) & - \lambda \bI_M + \bA_i^H \bA_i \bB_i^H \bB_i 
(\bB_i^H \bB_i + \lambda \bI_M)^{-1} 
\end{array} \right) 
\end{eqnarray}
which can be simplified to 
\begin{eqnarray} 
\begin{split} 
& \hspace{0.1in} \det ( \bC_1^T \bC_2 - \lambda \bI_{4M}) \\ 
& {\hspace{0.3in}}= 
\det(-\bB_i^H \bB_i - \lambda \bI_M) \det(-\lambda \bI_M) 
\det( - \lambda \bI_M + 
\bA_i^H \bA_i \bB_i^H \bB_i  (\bB_i^H \bB_i + \lambda \bI_M)^{-1}  ) 
\nonumber \\ 
& {\hspace{0.3in}}
\times \det( -\lambda \bI_M - \lambda \bA_i^{-H} 
(- \lambda \bI_M + \bA_i^H \bA_i \bB_i^H \bB_i  (\bB_i^H \bB_i + 
\lambda \bI_M)^{-1})^{-1} 
\bA_i^H (\bI_M - \bA_i \bA_i^H)  ). 
\end{split} 
\end{eqnarray}

Note that $\det(\bC_1^T \bC_2 - \lambda \bI_{4M}) = 0$ has $4M$ 
solutions for $\lambda$ with the solution from the first two terms being 
non-positive. Setting the fourth term to zero, and using the facts that 
$\det(\bI + \bC \bD) = \det( \bI + \bD \bC)$ and $\bI_M = \bA_i^H\bA_i + \bB_i^H \bB_i$, 
we see that $\lambda$ has to satisfy: 
\begin{eqnarray}
- 1 - \lambda_i( ( - \lambda \bI_M + \bA_i^H \bA_i (\bI_M - \bA_i^H \bA_i) 
(\bI_M + \lambda \bI_M -\bA_i^H \bA_i)^{-1} )^{-1} (\bI_M - \bA_i^H \bA_i) )
= 0. 
\end{eqnarray} 
After some straightforward simplifications, we can check that $\lambda$ 
is a solution to 
\begin{eqnarray}
\lambda_i(  \lambda (\bI_M -\bA_i^H \bA_i)^{-1} - \bA_i^H \bA_i 
(\bI_M + \lambda \bI_M - \bA_i^H \bA_i )^{-1} ) = 1. 
\label{plugger}
\end{eqnarray} 

Assume a singular value decomposition for $\bA_i$ and $\bB_i$ of 
the form: $\bA_i = \bU_{\bA} \bfLambda_{\bA}^{1/2} {\bf W}_{\bA}^H$ and $\bB_i = 
\bU_{\bB} \bfLambda_{\bB}^{1/2} {\bf W}_{\bB}^H$, respectively where $\bU_{\bA}, 
{\bf W}_{\bA}$ and ${\bf W}_{\bB}$ are $M \times M$ unitary matrices, and $\bU_{\bB}$ is 
an $(N_t - M) \times (N_t - M)$ unitary matrix. The full-rankness of $\bA_i$ 
means that the $M \times M$ diagonal matrix $\bfLambda_{\bA}$ is positive 
definite while the $(N_t - M) \times M$ matrix $\bfLambda_{\bB}$ has non-negative 
entries only along the leading diagonal. 
Since $\bA_i^H \bA_i + \bB_i^H \bB_i = \bI_M$, we have 
$\bI_M - \bfLambda_{\bA} = {\bf W}_{\bA}^H {\bf W}_{\bB} 
(\bfLambda_{\bB}^T \bfLambda_{\bB})^{1/2} {\bf W}_{\bB}^H {\bf W}_{\bA}$. Comparing the 
two sides, we see that ${\bf W}_{\bA} = {\bf W}_{\bB}$ (we set both to be ${\bf W}$) 
and 
$\bI_{M} - \bfLambda_{\bA} = (\bfLambda_{\bB}^T \bfLambda_{\bB})^{1/2}$. 
Note that since there are no constraints on$/$relationship between $\bU_{\bA}$ 
and $\bU_{\bB}$, the leading diagonal entries of $\bfLambda_{\bA}$ and 
$\bfLambda_{\bB}$ can be in any order. This is because either unitary matrix can 
be appropriately adjusted by a permutation matrix. 

Plugging in $\bA_i^H \bA_i = {\bf W} \bfLambda_{\bA} {\bf W}^H$ in~(\ref{plugger}), 
a routine computation 
yields $M$ solutions to $\lambda$ of the form: $\lambda^2 = 1 - \bfLambda_{\bA}(i)$. 
With the same form of $\bA_i^H \bA_i$, 
by setting the third term to zero, we obtain another 
$M$ solutions $\lambda = \sqrt{1 - \bfLambda_{\bA}(i)} \cdot \left( 
\frac { \sqrt{1 + 3 \bfLambda_{\bA}(i)}  - \sqrt{1 + \bfLambda_{\bA}(i)}  }{2} 
\right)$. Note that 
$\frac { \sqrt{1 + 3 \bfLambda_{\bA}(i)}  - \sqrt{1 + \bfLambda_{\bA}(i)}  }{2} < 1$ 
and hence, $\lambda_{\max}(\bC)$ is obtained by setting $i = M$ in the above 
solution which results in 
$\lambda_{\max}(\bC) = \sqrt{1 - \lambda_{\min}(\bA^H \bA)}$. 
Using this in (\ref{eqnb}), we get the expression for $\lambda_{\min}(\bA_i^H \bA_i)$. 
Furthermore, $\bI_M = \bA_i^H \bA_i + \bB_i^H \bB_i$ implies that 
\begin{eqnarray} 
1 = \lambda_{\max}(\bA_i^H \bA_i + \bB_i^H \bB_i) 
\geq \lambda_{\max}(\bA_i^H \bA_i) + \lambda_{\min}(\bB_i^H \bB_i) \geq 
\{ \lambda_{\max}(\bA_i^H \bA_i), \lambda_{\max}(\bB_i^H \bB_i) \}. 
\end{eqnarray} 
These are the constraints to be imposed on $\bfLambda_{\bA}$ to ensure that the 
scaling map preserves semiunitarity and reduces the minimum distance by $\alpha$.

If $M \leq (N_t - M)$, without loss in generality assume that the diagonal 
entries of $\bfLambda_{\bA}$ are in non-increasing order while those of 
$\bfLambda_{\bB}$ may be not. Given a choice of $\bfLambda_{\bA}$, the condition 
$\bI_{M} - \bfLambda_{\bA} = (\bfLambda_{\bB}^T \bfLambda_{\bB})^{1/2}$ can be 
met by choosing the principal $M \times M$ component of $\bfLambda_{\bB}$ to 
be $\left( \bI_M - \bfLambda_{\bA} \right)^{1/2}$. If $M > (N_t - M)$, assume 
that the diagonal entries of $\bfLambda_{\bB}$ are in non-increasing order 
while those of $\bfLambda_{\bA}$ may be not. Then, the condition $\bI_{M} - 
\bfLambda_{\bA} = (\bfLambda_{\bB}^T \bfLambda_{\bB})^{1/2}$ can be met if 
$2M - N_t$ entries of $\bfLambda_{\bA}$ are $1$. The additional 
constraint on the smallest diagonal entry (see discussion above) ensures distance 
scaling. 

To close the theorem, 
it is necessary to verify that $s(\bFa_1) = \bFa_1$. This can 
be done by checking that $\bfLambda_i$ can be computed in closed-form. For this, 
note that $\bfLambda_{\min} = 1$ and since $\bfLambda_{\max} \leq 1$, we have 
$\bfLambda_i = \bI_M$. From here, it can be checked that 
$\bB_i = \bO_{(N_t - M) \times M}$ and from (\ref{eqna}), we thus have 
$s(\bFa_1) = \bFa_1 \hsppp \bU_{\bA} \hsppp {\bf W}^H$. On the Grassmann manifold 
${\cal{G}}(N_t,M)$, multiplication by an $M \times M$ unitary matrix results in 
the same ``point.'' Thus $s(\bFa_1) = \bFa_1$ and the proof is complete. 
\end{proof}

Note that the choice of the scaling map is non-unique due to freedom in 
the choice of $\bU_{\bA}$, $\bU_{\bB}$ and ${\bf W}$ as well as the eigenvalues of 
$\bfLambda_i$ and ${\bf \Gamma}_i$. The case of ${\bf V}_i = {\bf V}_1$ is 
special where $\bfLambda_i$ turns 
out to be $\bI_M$. With almost any other choice of $\bFa_i$, these matrices are 
non-identity, in general. 
Besides these choices, non-uniqueness of the representative 
of $\bFa_1^{\nulla}$ also leads to non-uniqueness of the map. 

\bibliographystyle{IEEEbib}
\bibliography{newrefsx}

\end{document}